\documentclass[conference]{IEEEtran}

\ifCLASSINFOpdf
\else
\fi

\usepackage{mathptmx}  %

\usepackage[T1]{fontenc}
\usepackage[utf8]{inputenc}
\usepackage{pslatex}
\usepackage{graphicx}
\usepackage{subfig}
\usepackage{booktabs}
\usepackage{threeparttable}
\usepackage{array}
\usepackage{amssymb}
\usepackage{enumitem}
\usepackage{pifont}
\usepackage{multirow}
\usepackage{makecell}
\usepackage{tablefootnote}
\usepackage{titlesec}
\usepackage{threeparttable}
\usepackage{siunitx}
\usepackage{tabularx}
\usepackage[normalem]{ulem}
\usepackage[framemethod=tikz]{mdframed}

\usepackage[kerning,spacing]{microtype}

\usepackage{cite}
\usepackage{breakurl}
\usepackage{url}
\usepackage{xcolor}
\usepackage[]{hyperref}
\hypersetup{
  colorlinks,
  linkcolor={green!80!black},
  citecolor={red!70!black},
  urlcolor={blue!70!black}
}
\usepackage{amsmath}
\usepackage{cleveref}
\usepackage{placeins}

\Crefname{figure}{Fig.}{Figs.}
\Crefname{table}{Table}{Tables}
\crefname{appendix}{Appendix}{Appendices}
\Crefname{appendix}{Appendix}{Appendices}
\newcolumntype{M}[1]{>{\centering\arraybackslash}m{#1}} %
\newcolumntype{R}[1]{>{\raggedleft\arraybackslash}p{#1}} %
\newcolumntype{L}[1]{>{\raggedright\arraybackslash}p{#1}} %

\newcommand{\nop}[1]{}
\newcommand\featuretext[1]{
  \llap{\vrule width.35pt height2pt depth2.5pt\kern1pt}%
  \rlap{\rotatebox{35}{\textbf{#1}}}%
}

\newtheorem{theorem}{Theorem}
\newtheorem{corollary}{Corollary}[theorem]

\usepackage{xspace}
\newcommand{\sys}{\textsc{I2perception}\xspace}

\newmdenv[
  backgroundcolor=blue!5,      %
  linecolor=blue!60!black,     %
  linewidth=0.8pt,              %
  roundcorner=2pt,              %
  innerleftmargin=6pt,          %
  innerrightmargin=6pt,         %
  innertopmargin=4pt,           %
  innerbottommargin=4pt,        %
  skipabove=6pt,                %
  skipbelow=6pt                 %
]{importantblock}

\makeatletter
\g@addto@macro\normalsize{%
  \setlength\abovedisplayskip{4pt}%
  \setlength\belowdisplayskip{4pt}%
  \setlength\abovedisplayshortskip{2pt}%
  \setlength\belowdisplayshortskip{2pt}%
}
\makeatother

\begin{document}

\renewcommand{\footnoterule}{%
  \kern -3pt
  \hrule width 3.5in height 0.8pt
  \kern 2.6pt
}

\title{\Large \bf Time will Tell: Large-scale De-anonymization of Hidden I2P Services via Live Behavior Alignment (Extended Version)}

\author{\IEEEauthorblockN{Hongze Wang\IEEEauthorrefmark{2},
Zhen Ling\IEEEauthorrefmark{2}\IEEEauthorrefmark{1} \thanks{* Corresponding author: Prof. Zhen Ling of Southeast University, China.},
Xiangyu Xu\IEEEauthorrefmark{2},
Yumingzhi Pan\IEEEauthorrefmark{2},
Guangchi Liu\IEEEauthorrefmark{2}, 
Junzhou Luo\IEEEauthorrefmark{2}\IEEEauthorrefmark{3} and
Xinwen Fu\IEEEauthorrefmark{4}}
\IEEEauthorblockA{\IEEEauthorrefmark{2}Southeast University,
Email: \{wanghongze, zhenling, xy-xu, pymz, gc-liu, jluo\}@seu.edu.cn}
\IEEEauthorblockA{\IEEEauthorrefmark{3}Fuyao University of Science and Technology}
\IEEEauthorblockA{\IEEEauthorrefmark{4}University of Massachusetts Lowell, Email: xinwen\_fu@uml.edu}
} %

\IEEEoverridecommandlockouts
\makeatletter\def\@IEEEpubidpullup{6.5\baselineskip}\makeatother

\maketitle

\begin{abstract}
I2P (Invisible Internet Project) is a popular anonymous communication network. While existing de-anonymization methods for I2P focus on identifying potential traffic patterns of target hidden services among extensive network traffic, they often fail to scale effectively across the large and diverse I2P network, which consists of numerous routers. In this paper, we introduce \sys a low-cost approach revealing the IP addresses of I2P hidden services. In \sys, attackers deploy floodfill routers to passively monitor I2P routers and collect their \textit{RouterInfo}. We analyze the router information publication mechanism to accurately identify routers' join (i.e. on) and leave (i.e. off) behaviors, enabling fine-grained live behavior inference across the I2P network. Active probing is used to obtain the live behavior (i.e., on-off patterns) of a target hidden service hosted on one of the I2P routers. By correlating the live behaviors of the target hidden service and I2P routers over time, we narrow down the set of routers matching the hidden service’s behavior, revealing the hidden service’s true network identity for de-anonymization. Through the deployment of only 15 floodfill routers over the course of eight months, we validate the precision and effectiveness of our approach with extensive real-world experiments. Our results show that \sys successfully de-anonymizes all controlled hidden services.
\end{abstract}

\section{Introduction}
\label{sec:intro}

I2P (Invisible Internet Project) is a widely used anonymous communication network, supporting activities such as website browsing, email, file sharing, Internet Relay Chat (IRC), and cryptocurrency access, with around 45,000 active routers and over 15,000 daily users and services \cite{i2pmetrics}. As a decentralized peer-to-peer system, I2P enables anonymous data transmission through routers that voluntarily share bandwidth and storage. A subset of routers, referred to as {\em floodfill} routers, maintain a distributed database containing access information for all routers and hidden services on I2P. Like Tor \cite{dingledine2004tor}, I2P uses multilayer encryption and multihop tunnels to secure communication, ensuring that no single router can correlate client-service interactions or identify the hosting router of a specific hidden service, for the sake of privacy and secure communication. \looseness=-1

To assess the security and limitations of anonymous networks, a large volume of prior work has investigated deanonymizing methods. These efforts have primarily concentrated on the Tor network, one of the most prevalent anonymous communication systems, and are mainly based on the traffic correlation approaches \cite{ling_new_2009,murdoch_low-cost_2005,wang_network_2007,mittal_stealthy_2011,biryukov_trawling_2013,houmansadr_swirl_2011,borisov_denial_2007,chakravarty_traffic_2010,evans_practical_2009,wacek_empirical_2013,edman_as-awareness_2009,matic_caronte_2015,kwon_circuit_2015,rochet_claps_2020,sun_counter-raptor_2017,oh_deepcoffea_2022,nasr_deepcorr_2018,berthold_dummy_2003,lopes_flow_2024,overlier_locating_2006,bauer_low-resource_2007,nithyanand_measuring_2015,backes_nothing_2014,sun_raptor_2015,shmatikov_timing_2006,johnson_users_2013,rochet_waterfilling_2017, Zhang_Paxson_2000,Blum_Song_Venkataraman_2004,mathewson2004practical,danezis_two-sided_2007,murdoch_sampled_2007}. Such approaches typically require the attacker to first use techniques like DoS attacks to occupy key positions in the target hidden service's circuit \cite{jansen2019point,jansen2014sniper,barbera2013cellflood}, in order to conduct traffic correlation. However, while I2P employs a multi-hop tunnel architecture comparable to Tor's circuit for achieving anonymity, it adopts a different approach from Tor for selecting participants for a hidden service's tunnel. Given the current scale of the I2P network, it is exceedingly difficult for an attacker to successfully control the routers participating in a target hidden service's tunnels, thereby making traffic correlation unachievable. 

Although I2P's network communication mechanisms and scale seemingly allow a hidden service to hide well within the `crowd', we demonstrate that the online and offline patterns of a hidden service, referred to as its live behavior, can still serve as a powerful side-channel for compromising its anonymity. Our approach relies on two key observations. First, an I2P hidden service is hosted on an I2P router, resulting in its live behavior to closely align with that of its hosting router. Second, user behavior in peer-to-peer (P2P) systems, including I2P, is inherently diverse \cite{gummadi2003measurement,rhea2004handling,saroiu2003measuring,sen2002analyzing,sripanidkulchai2004analysis}, causing significant variations in router behavior due to the heterogeneity of user activity. Therefore, by identifying the router whose live behavior most closely aligns with that of a target hidden service, we can accurately infer the location of the hosting router and de-anonymize the hidden service.
\looseness=-1

However, the practical implementation of the aforementioned deanonymization method presents significant challenges. First, achieving large-scale, stealthy, and fine-grained monitoring of I2P routers’ live behavior simultaneously is highly difficult. Active probing of approximately 45,000 routers daily is impractical due to excessive overhead, the risk of exposing probing intentions, and potential degradation of network performance.
Passive monitoring faces grand challenges too, as only \textit{RouterInfo} (which contains the router's contact information and is periodically published to floodfill routers) is passively obtainable, and I2P employs obfuscation mechanisms to prevent \textit{RouterInfo} from revealing detailed behavior information.
Second, even if the router's live behavior could be inferred from \textit{RouterInfo}, the large scale of I2P makes it exceedingly difficult to collect complete data in a cost-effective manner. Accurately inferring a router's live behavior under incomplete data conditions poses a significant challenge. \looseness=-1

In this paper, we introduce \sys, a large-scale deanonymization approach designed to reveal the real IP address of a target I2P hidden service. First, we leverage only 15 floodfill routers to passively collect \textit{RouterInfo} at low cost from all I2P routers to create a \textit{RouterInfo} trace for each router.
Using these traces, we exploit the periodic \textit{RouterInfo} publication patterns across different I2P implementations to achieve coarse-grained online session inference of each router. 
We then investigate the \textit{RouterInfo} publication mechanisms during the join and leave periods of routers to precisely identify their join and leave behaviors, achieving fine-grained router behavior inference. 
To mitigate the impact of incomplete data due to our low-cost data collection strategy, we propose a data-recovery-based online session complement method. 
Next, we apply active probing to infer the live behavior of the target hidden service.
Finally, we introduce a live behavior correlation method to measure the similarity between the inferred live behaviors of routers and the target hidden service, allowing us to identify the host router of the hidden service. \looseness=-1

Our major contributions can be summarized as follows:
\begin{itemize}
    \item We propose a large-scale and long-lasting I2P hidden server deanonymization framework, \sys, based on the similarity of the live behavior of the target hidden service and I2P routers, which can be used to deanonymize any target hidden service in the I2P network within a proper monitoring window.

    \item We design methods that can accurately identify the join and leave behaviors of routers by delving into the \textit{RouterInfo} publication mechanism during the startup and offline periods.
    We introduce an online session recovery method to mitigate the impact of incomplete data caused by the low-cost data collection strategy, achieving highly accurate inferred live behaviors for all routers in the I2P network.

    \item To demonstrate the feasibility and effectiveness of \sys, we conduct extensive real-world experiments by deploying 15 floodfill routers and 10 I2P routers hosting hidden services over a period of approximately eight months. The accuracy of deanonymizing our controlled hidden services approaches nearly 100\%.

\end{itemize}

The key novelty of this work is that we exploit a user’s characteristic on–off behavior when operating an I2P hidden service and introduce new measurement techniques that accurately capture that pattern to deanonymize the service. The scientifically generalizable takeaway is that user behavior can serve as a measurable side channel that leaks identity. 

{\bf Responsible Disclosure}. Upon discovering the vulnerabilities, we disclosed our findings to the I2P maintainers to support timely mitigation. Our work was acknowledged by the I2P project, and as of this submission, the reported issues have been addressed \cite{i2pRepairedRelease}. \looseness=-1

\section{Background}
\label{sec:background}

In this section, we first provide an overview of the I2P network structure, and then present its communication mechanism. \looseness=-1

\subsection{Architecture of I2P}
\label{subsec:I2P_architecture}

\begin{figure}[thb!]
    \centering
    \includegraphics[width=0.92\linewidth]{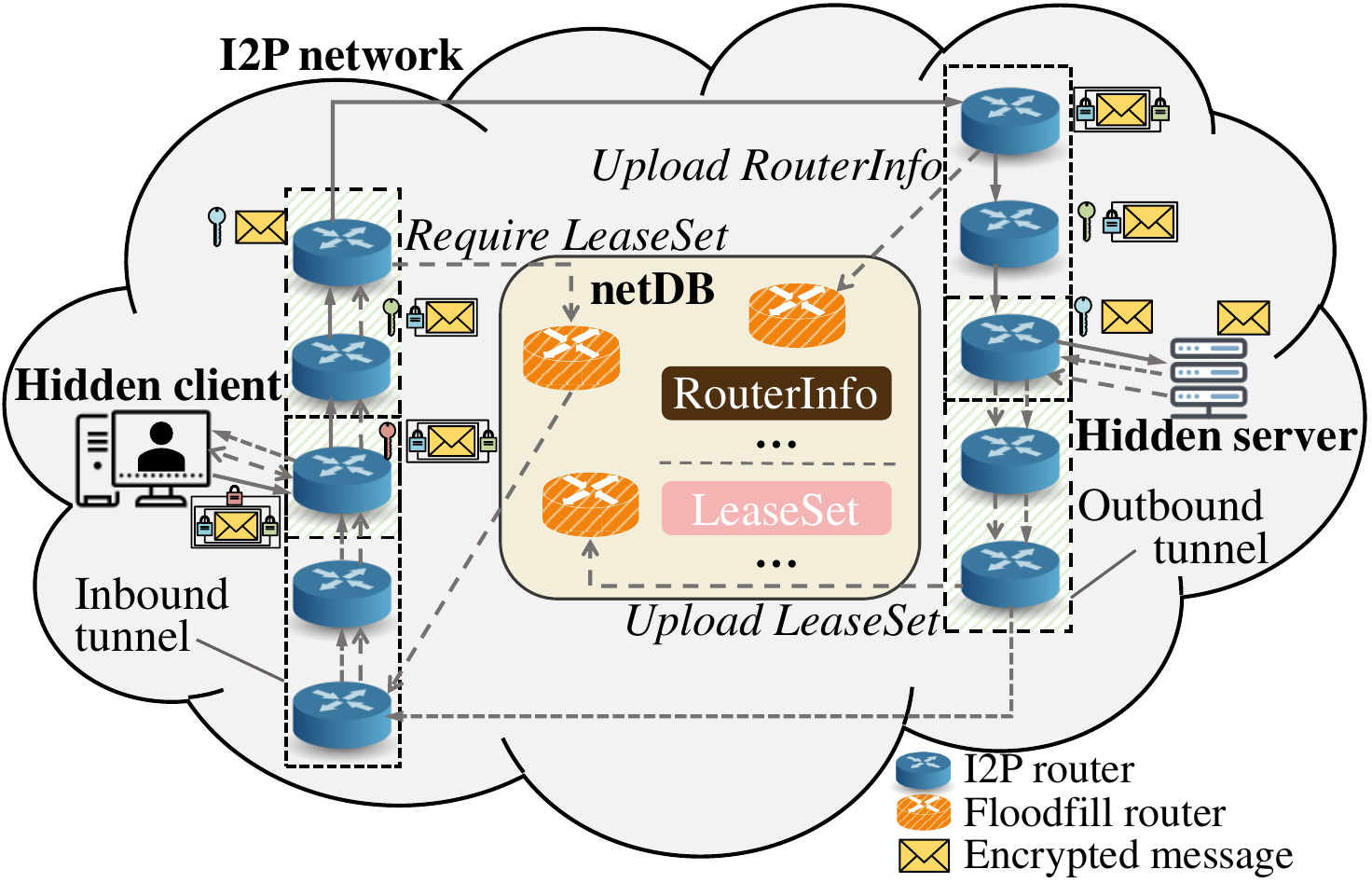}
    \caption{I2P Architecture.}
    \label{fig:I2P_arch}
\vspace{-4mm}
\end{figure}

The Invisible Internet Project (I2P) is a decentralized peer-to-peer communication system enabling anonymous data transmission. As shown in \Cref{fig:I2P_arch}, the I2P network comprises four key components: I2P routers, I2P clients, hidden services, and netDB \cite{Official}. \looseness=-1

\textbf{I2P router}. I2P routers form the backbone of I2P's communication infrastructure. 
Upon installation, router functionality is enabled by default. During its initial startup, the router contacts specific bootstrap sites to discover other routers, thereby joining the I2P network and relaying data for other participants. Routers can operate in either \textit{floodfill} mode or \textit{non-floodfill} mode. While non-floodfill routers primarily relay data, floodfill routers also gather and maintain information about all routers in the I2P network, including both \textit{floodfill} and \textit{non-floodfill} routers. \looseness=-1

\textbf{I2P client}. The I2P software includes several default application clients, such as an IRC client and a BitTorrent client. The software also supports an HTTP proxy, allowing various clients to connect to the I2P network through the router. A router that hosts a client, referred to as a \textit{host router}, relays client data via multiple two-hop, unidirectional inbound and outbound tunnels.\looseness=-1

\textbf{I2P hidden service}. I2P provides a web service for users to deploy hidden web services on the network. Additionally, an I2P router supports a simple anonymous messaging (SAM) protocol, enabling various services to interact with the host router. Like an I2P client, a hidden service uses the host router to establish inbound and outbound tunnels to relay its data. \looseness=-1

\textbf{netDB}. The netDB is a distributed database composed of all floodfill routers, maintaining information about I2P routers and hidden services. Floodfill routers are organized using a distributed hash table (DHT) based on the Kademlia algorithm \cite{kademlia}. The netDB stores two types of data: \textit{RouterInfo} and \textit{LeaseSet}. \textit{RouterInfo} contains router details, such as IP address, port, and public key. An I2P router periodically selects a floodfill router to upload its \textit{RouterInfo}, with the floodfill router returning an acknowledgment message confirming the upload. A hidden service or client records its details, including a public key and the entry router of its inbound tunnel in the \textit{LeaseSet} data. Hidden services upload their \textit{LeaseSet} to the netDB, while clients send their \textit{LeaseSet} to a target hidden service via the outbound tunnel when interacting with the service. A \textit{LeaseSet} expires based on the service's configuration, while \textit{RouterInfo}, expires after one hour unless updated. \looseness=-1

\subsection{Garlic Routing}
\label{subsec:garlic}
I2P uses garlic routing to facilitate anonymous communication between hidden services and clients. Routers obtain \textit{RouterInfos} from the netDB (floodfill routers) to discover other routers. A hidden service or client relies on its host router to select suitable routers and establish paired, unidirectional, two-hop inbound and outbound tunnels for data transmission, as shown in \Cref{fig:I2P_arch}. Each tunnel consists of three I2P routers, including the host router, and the data transmitted through these tunnels are called garlic messages. To connect to a hidden service, a client derives the service's I2P address and queries the netDB to retrieve its \textit{LeaseSet}, which contains the entry router for the service's inbound tunnel. The client then selects an outbound tunnel to establish the connection, sending its \textit{LeaseSet} to the hidden service. By learning the entry router for the client's inbound tunnel from the \textit{LeaseSet}, the service initiates a connection to the client's inbound tunnel through its outbound tunnel. Bidirectional communication is achieved through the concatenated four-hop tunnels. Each router in these tunnels can only identify the IP addresses of its immediate predecessor and successor, making it difficult to correlate client-server communication. \looseness=-1

\begin{figure}[t]
    \centering
    \includegraphics[width=0.98\linewidth]{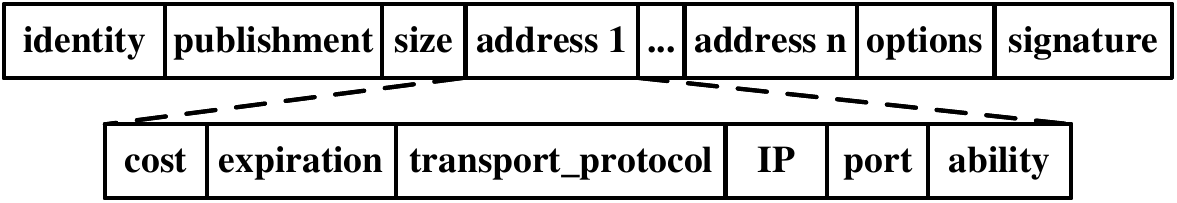}
    \caption[short]{Format of RouterInfo data.}
    \label{fig:ri_arch}
\vspace{-5mm}
\end{figure}

\subsection{RouterInfo}
\label{subsec:RouterInfo}

\textbf{Structure.} \Cref{fig:ri_arch} illustrates the structure of the \textit{RouterInfo}. The first field, \textit{Identity}, uniquely identifies the router within the I2P network and is generated upon the router's initial launch. The \textit{publishment} field records the creation time of the \textit{RouterInfo}, while the \textit{size} field indicates the number of associated \textit{RouterAddress} entries. Each \textit{RouterAddress} contains essential connection details, such as \textit{cost}, \textit{expiration} (deprecated), I2P transport protocol (NTCP or SSU), IP address, port, and an \textit{ability} field. The \textit{cost} value is assigned by the router, where a lower cost increases the likelihood of being chosen for connections. The \textit{ability} field indicates whether the router offers a \textit{peer testing} service (with a `B' flag) or an \textit{introducing} service (with a `C' flag). The `B' flag means the router can test reachability, while the `C' flag indicates it can function as a reverse proxy for NAT-bound routers. The final two fields, \textit{signature} and \textit{options}, ensure integrity and store the router's profile, including operational mode, bandwidth, congestion level, port reachability, and version number. Floodfill routers are marked with an `f' flag in the \textit{options} field, distinguishing them from non-floodfill routers. \looseness=-1 

\textbf{Propagation.} I2P implements a specific mechanism for uploading and propagating \textit{RouterInfos}. Each router selects a target floodfill router to upload its \textit{RouterInfo}, based on a parameter known as the \textit{Routingkey}, which is computed using the \textit{SHA256} hash algorithm as follows:
\begin{equation}
    \label{eq:routing_key}
    {Routingkey}={Hash}({Hash}(identifier) + date)
\end{equation}
Here, \textit{date} denotes the formatted timestamp of the \textit{RouterInfo} publication, and \textit{identifier} refers to the router’s unique ID within the I2P network. The router then selects the floodfill router whose identifier has the smallest \textit{XOR} distance to its own \textit{Routingkey} \cite{kademlia}. However, since the router lacks global knowledge of all floodfill routers in the network, there may exist others with identifiers even closer to its \textit{Routingkey}. To address this, each floodfill router is also responsible for further propagating \textit{RouterInfos}. After extracting the originating router’s identifier and publication time, a floodfill router recomputes the corresponding \textit{Routingkey} and forwards the \textit{RouterInfo} to three additional floodfill routers from its known set, based on the same selection rule. Forwarding is skipped if no suitable floodfill routers are available or if the \textit{RouterInfo} has already been processed by the router. This propagation mechanism ensures that each \textit{RouterInfo} data could be received by numerous floodfill routers upon publication. \looseness=-1

\section{Deanonymizing I2P via Live Behavior}
\label{sec:prob&obs}

\subsection{Problem Statement}
\label{subsec:problem}

Our objective is to stealthily uncover the real IP addresses of target I2P hidden services. Specifically, we aim to identify the I2P router hosting the hidden service. However, this task can be challenging. As depicted in \S\ref{subsec:garlic}, the multi-hop \textit{garlic routing} ensures that both the client and the server only know the IP addresses of their respective tunnel entry routers, without revealing each other's actual IP addresses. Additionally, each participant router within these tunnels is only aware of the IP addresses of its predecessor and successor. This design prevents routers from not only correlating the communication relationship between the client and the hidden services but also identifying whether a specific hidden service is hosted on an I2P router.\looseness=-1

\subsection{Threat Model} 
We make three key assumptions for our method. First, an attacker can deploy several floodfill routers to join the I2P network and collect \textit{RouterInfo} data from other routers. This is feasible due to I2P's open and decentralized nature, which allows anyone to deploy routers and participate. By modifying the configuration, the attacker can enable floodfill mode on these routers to monitor routers' live behavior using their uploaded \textit{RouterInfos}. Second, the attacker targets a specific hidden service for deanonymization, which is accessible without restrictions. The target server cannot distinguish between the attacker’s access and that of other users, preventing it from revealing the attacker's IP address. This allows the attacker to observe the service's live behavior.
Third, the target hidden service is not hosted on a router in \textit{hidden mode}. Collecting hidden routers' \textit{RouterInfo} is extremely challenging, making them nearly impossible to monitor. However, this assumption is reasonable since routers operate in \textit{non-hidden} mode by default and, according to our long-term measurements of the I2P network, we believe that hidden routers constitute a very small fraction of the network. The difficulty in obtaining their \textit{RouterInfo} arises for two reasons: (1) floodfill routers do not broadcast the \textit{RouterInfo} of hidden routers; and (2) hidden routers reject all inbound connections and can only be detected by routers, including floodfill routers, to which they initiate outbound connections. Over the monitor of one year, we deployed 15 floodfill routers, added 5 more for two months, and operated an additional 10 non-floodfill routers for eight months. Throughout this period, none of our controlled routers were contacted by any hidden routers. 
\looseness=-1

\subsection{Basic Idea}
\label{subsec:observation}

We leverage two fundamental observations that arise from the characteristics of I2P's design to de-anonymize a target hidden service. First, since each hidden service is hosted on an I2P router, its live behavior mirrors that of the hosting router. This means that by identifying the router's live behavior, we can infer the hidden service’s behavior. In addition, we observe that I2P routers periodically upload their \textit{RouterInfo} to the netDB, a regular pattern that can potentially be used to infer their live behaviors.

Second, the live behaviors of I2P routers exhibit significant variation. Previous studies \cite{hoang2018IMC} show that only 31.15\% of I2P routers remain active for more than 30 days, with many exhibiting intermittent availability. Other research \cite{gummadi2003measurement,rhea2004handling,sen2002analyzing} on P2P networks indicates that user behavior is inherently diverse, with substantial behavioral differences among users. This diversity in router behaviors further reduces the anonymity set — the group of I2P routers with similar behaviors — making it easier to distinguish individual routers. \looseness=-1

By combining these observations, an attacker can monitor the live behaviors of I2P routers and correlate them with the live behavior of the target hidden service. Due to behavioral divergence among routers, the overlap between their live behaviors naturally decreases over time. As the monitoring period extends, the number of routers whose behavior matches the target service gradually decreases. This allows the attacker to progressively narrow down the set of potential hosting routers, ultimately achieving deanonymization. \looseness=-1

\subsection{Challenges and Solutions}
\label{sec:challenges}

While deploying the de-anonymization approach is theoretically straightforward, several practical challenges arise during its implementation. We address these challenges as  follows.

\textbf{(C-\uppercase\expandafter{\romannumeral1}) Large-scale Analysis of Routers' Fine-grained Live Behaviors.} A typical method for inferring the fine-grained live behavior of an I2P router involves periodically and actively probing its availability. An attacker can leverage the router's IP address and port information from the \textit{RouterInfo} to perform port scanning. However, this approach introduces significant overhead, especially when monitoring a large number of I2P routers simultaneously (approximately 45,000 active routers each day \cite{i2pmetrics}). Furthermore, active probing risks exposing the attacker's intentions, potentially alerting the targets and prompting evasive actions. Consequently, passive monitoring approaches, which can unobtrusively monitor the live behavior of a large number of routers, are crucial for large-scale analysis.

However, passive monitoring poses specific challenges due to limited data availability for inferring router behavior. While I2P routers periodically upload \textit{RouterInfo} to the netDB---an activity that can be leveraged to infer coarse-grained live behavior---the system intentionally introduces randomness into the intervals between successive routine uploads to obfuscate predictability. This randomness undermines the reliability of coarse-grained inference methods, increasing their susceptibility to errors, particularly when routers experience short-term offline periods. Consequently, fine-grained identification of online sessions, which is capable of precisely determining router online and offline times, remains a nontrivial task.

\begin{figure*}[!th]
    \centering
    \includegraphics[width=0.98\textwidth]{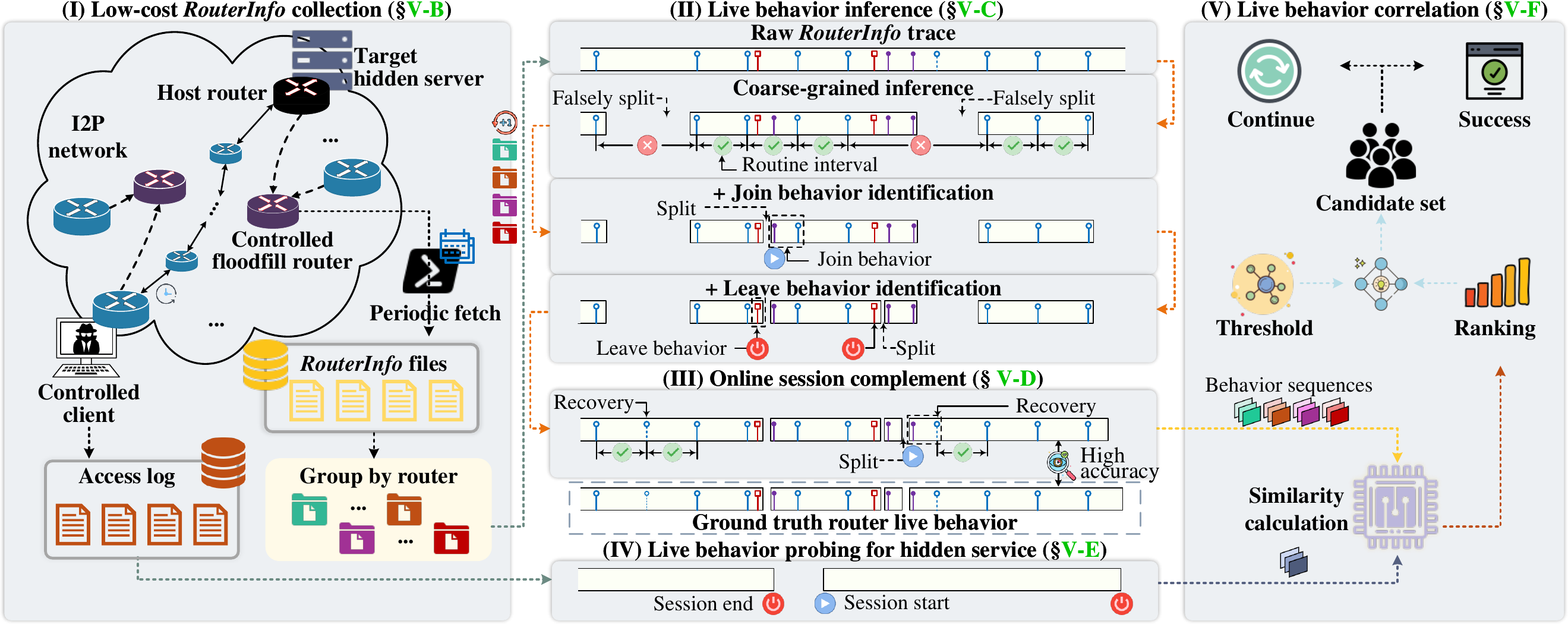}
    \caption[short]{\sys Overview.}
    \label{fig:overview}
\vspace{-5mm}
\end{figure*}

\textbf{(S-\uppercase\expandafter{\romannumeral1}) RouterInfo-publication-based Live Behavior Inference (\S\ref{subsec:time_state}).} Through a detailed investigation of the \textit{RouterInfo} publication mechanism, we propose a passive, low-cost, large-scale monitoring method for identifying fine-grained live behavior. Regardless of the I2P implementation (Java or C++), an online I2P router is designed to perform multiple tasks to generate and upload its \textit{RouterInfo} to the netDB (i.e., floodfill routers), thereby informing other routers in the network of its current status, such as reachability and load. We focus on the periodically executed tasks and analyze the characteristics of the generated \textit{RouterInfo} to identify which publications are associated with these tasks. By only passively deploying 15 floodfill routers, we can infer the implementation of the I2P router based on the periodic update intervals and further identify gaps between routine publications that are longer than expected, and interpret them as offline periods. To achieve fine-grained identification of temporary offline behavior, we find that I2P routers display distinctive \textit{RouterInfo} publishing patterns when going online and offline, allowing us to determine the exact start and end times of each online session. \looseness=-1

\textbf{(C-\uppercase\expandafter{\romannumeral2}) Accuracy of Low-cost Live Behavior Identification.} Our live behavior inference method for I2P routers relies heavily on \textit{RouterInfos} collected by only 15 controlled floodfill routers. However, gathering complete \textit{RouterInfos} from a large number of routers may require deploying hundreds or even thousands of additional floodfill routers, as approximately 4,000 already operate daily, servicing around 45,000 active routers \cite{i2pmetrics}. While deploying sufficient floodfill routers can capture \textit{RouterInfos} to accurately infer online sessions, this approach introduces substantial overhead and may compromise stealth. To reduce costs, the attacker may limit the number of floodfill routers, resulting in incomplete \textit{RouterInfo} collection, which complicates accurate online session identification. \looseness=-1

Additionally, while the \textit{RouterInfo} propagation mechanism (\S\ref{subsec:RouterInfo}) enables the attacker to collect most network-wide data with a limited number of floodfill routers, network churn--—the continuous arrival and departure of thousands of routers—--can cause \textit{RouterInfos} to be uploaded to different floodfill routers or propagated via different paths over time. This variability leads to partial \textit{RouterInfo} collection for each I2P router, reducing the accuracy of live behavior inference and causing mismatches when comparing router behaviors to that of the target hidden service. Mitigating the impact of incomplete \textit{RouterInfo} data on inference accuracy remains a significant challenge.\looseness=-1

\textbf{(S-\uppercase\expandafter{\romannumeral2}) Data-recovery-based Session Complement (\S\ref{subsec:complement}).} We propose a router behavior complement method to mitigate the impact of incomplete \textit{RouterInfo} data. Each online session is expected to include explicit start-up and offline behaviors. For incomplete sessions, we first determine whether the issue arises from a single session being erroneously split into two due to missing \textit{RouterInfo}, or from absent startup or shutdown indicators. Based on this determination, we adopt a context-aware strategy to insert the missing \textit{RouterInfos} at suitable points, allowing for accurate reconstruction of the session.

\section{Design of \sys}
\label{sec:overview}
In this section, we first present the workflow of \sys and then introduce its five phases in detail. A theoretical analysis is provided at the end to show why the target hidden service can be uniquely identified.
\looseness=-1

\subsection{Workflow}
Figure~\ref{fig:overview} illustrates the overall workflow of \sys, which consists of five sequential phases.
(i) {\em Low-cost RouterInfo collection} (\S\ref{sec:data_collect}).
We passively collect RouterInfo from all observable I2P routers using a small number of controlled floodfill routers, constructing a RouterInfo trace for each router. 
(ii) {\em Fine-grained router live behavior inference} (\S\ref{subsec:time_state}). Using these RouterInfo traces, \sys infers each router’s online sessions. We first derive coarse-grained sessions based on periodic RouterInfo publication, and then identify precise join and leave behaviors to obtain fine-grained online/offline intervals. 
(iii) {\em Online session complement} (\S\ref{subsec:complement}). Since low-cost data collection may miss certain RouterInfo, \sys applies a data-recovery–based complement mechanism to reconstruct incomplete sessions and restore missing join or leave behaviors. 
(iv) {\em Live behavior probing for the target hidden service} (\S\ref{sec:server_state}). In parallel, \sys actively probes the target hidden service to obtain its live behavior sequence, which reflects the availability of its hosting router. 
(v) {\em Live behavior correlation} (\S\ref{subsec:similar_alg}). Finally, \sys computes the similarity between the hidden service’s live behavior and that of all routers. The router with the highest behavioral consistency over time is identified as the hosting router. \looseness=-1

\subsection{Low-cost RouterInfo Collection}
\label{sec:data_collect}

We collect \textit{RouterInfos} from I2P routers by deploying only 15 controlled floodfill routers to participate in the netDB. Since I2P routers actively upload their latest \textit{RouterInfo} to the netDB, we can passively gather this data from the controlled floodfill routers. Upon receiving \textit{RouterInfos}, the routers extract and store the data fields into our database, allowing us to compile a series of \textit{RouterInfo} traces for each router.
We demonstrate the effectiveness of the low-cost collection approach using only 15 floodfill routers in Appendix~\ref{subsec:floodfill_num}. \looseness=-1

\subsection{Live Behavior Inference}
\label{subsec:time_state}

\begin{table*}[tp]
    \centering
    \caption{Inherent mechanisms and exploitation approaches of each live behavior inference framework component.}
    \label{tab:comparison}
    
    \begin{tabular}{M{1.6cm} | M{0.7cm} | L{6.5cm} | L{7.5cm}}
        \toprule[1.5pt]
        \textbf{Framework component} & \textbf{Router} & \textbf{Inherent mechanism} & \textbf{Exploitation approach} \\
        \midrule[1.5pt]
        
        \multirow{2}{=}{\textbf{Online session inference}}
        & Java  
        & Evaluate router status every 8–10 minutes, and forcibly publish a routine \textit{RouterInfo} after every four evaluations.
        & Use a 32--43-minute interval to identify routine \textit{RouterInfos} and divide sessions by how continuously they appear. \\

        \cmidrule{2-4}
        
        & C++ 
        & Evaluate congestion level every 12 minutes; update and republish a \textit{RouterInfo} if the congestion flag changes.
        & Treat \textit{RouterInfos} with congestion-flag changes as routine. Split online sessions when two consecutive routine \textit{RouterInfos} are not separated by a multiple of 12 minutes.\\
        
        \midrule

        \multirow{2}{=}{\textbf{Join behavior identification}}
        & Java 
        & Upon startup, a router publishes an initial \textit{RouterInfo}, which has a distinctive time gap (i.e., startup interval) from the first routine \textit{RouterInfo}.
        & Using the startup interval, trace back from the first routine \textit{RouterInfo} in an online session to find the initial \textit{RouterInfo} and determine the join time. \\

        \cmidrule{2-4}
        
        & C++ 
        & At startup, the router initializes the congestion timer and issues its first \textit{RouterInfo} after 0.5 seconds.
        & Trace back from the first routine \textit{RouterInfo} to one published $12 min \times n - 0.5 s$ earlier, thereby identifying the router's join time. \\
        
        \midrule

        \multirow{2}{=}{\textbf{Leave behavior identification}}
        & Java 
        & Before going offline, a floodfill router publishes a \textit{RouterInfo} announcing it is no longer floodfill. 
        & Determine the router’s leave time by detecting when its operation mode changes. (floodfill only)\\

        \cmidrule{2-4}
        
        & C++ 
        & Before going offline, the router publishes a \textit{RouterInfo} with an `E' congestion flag to prevent new connections.
        & Find the \textit{RouterInfo} with the highest congestion level; its publication time is the router’s leave time. \\
        
        \bottomrule[1.5pt]
    \end{tabular}
    \vspace{-4mm}
\end{table*}

Our live behavior inference framework consists of three components: (i) \textbf{Online session inference}: We utilize the periodic status-maintenance mechanism of I2P routers, through which they generate and publish routine \textit{RouterInfo}. By extracting all routine \textit{RouterInfos} from the raw trace and analyzing the time intervals between them, we determine whether they are published within the same online session, enabling a coarse identification of each router’s online and offline periods. (ii) \textbf{Join behavior identification}: We exploit the distinctive \textit{RouterInfo} publication pattern exhibited by I2P routers during startup to identify join behavior at the beginning of each online session, thereby determining the router’s precise join time. (iii) \textbf{Leave behavior identification}: For routers that publish a \textit{RouterInfo} as part of their shutdown procedures, we identify these leave behavior \textit{RouterInfos} at the end of each online session to determine the router’s exact leave time. \Cref{tab:comparison} summarizes the inherent \textit{RouterInfo} management and publication mechanisms leveraged by each component, as well as how these mechanisms are exploited across different I2P router implementations. \looseness=-1

For clarity of exposition, we describe the fine-grained inference techniques using Java-based I2P routers below, as Java is the original implementation of I2P and most faithfully reflects the core design principles of the system. The corresponding procedures for C++-based routers, including coarse-grained inference, join behavior identification, and leave behavior identification, are presented in Appendix~\ref{subsubsec:cpp_code_analysis}, while the approach for distinguishing the two implementations of routers is present in Appendix~\ref{subsubsec:versions}. The overall analytical structure remains the same, with differences arising primarily from implementation-specific publication schedules and event-driven behaviors. \looseness=-1

\textbf{Coarse-grained online session inference for Java-based routers.} Java-based routers periodically execute an \textbf{update task} that checks their status (e.g., network congestion) and publishes a new \textit{RouterInfo} if any status change is detected. The update interval, denoted as $D_c$, typically ranges from 8 to 10.5 minutes, calculated as follows:
\vspace{-1mm}
\begin{equation}
    \label{eq:check_interval}
    D_c=\left(T\times \frac{3}{4} + random\left(S\right)\right) \div 4,
\end{equation}
where $T$ is a constant (default value of 43) and $S$ is a random value between 0 and 10, to obfuscate the update time. Additionally, the execution of an update task is delayed until at least nine minutes have elapsed since the previous successful \textit{RouterInfo} publication. Regardless of status changes, a router publishes a \textit{RouterInfo} every fourth update task (approximately every 32 to 43 minutes, i.e., $4 \times D_c$). We refer to these periodic publication as \textbf{routine publish tasks}, and the data generated by them as \textbf{routine \textit{RouterInfos}}. \looseness=-1

We identify a Java-based router's online sessions by analyzing its routine \textit{RouterInfos}, distinguishable by specific publication intervals. If the interval between two \textit{RouterInfo} publications equals four update intervals ($4 \times D_c$), such pairs are classified as routine \textit{RouterInfos}. Starting from the first \textit{RouterInfo} in a trace, we check if a subsequent publication appears within 32 to 43 minutes. If not, the first data point is not considered routine. This process continues until a pair of routine \textit{RouterInfos} is found. Once identified, each routine \textit{RouterInfo} serves as a baseline to locate the next, repeating the process until no further routine \textit{RouterInfo} is found. All identified routine \textit{RouterInfos} and the data between them are grouped into a single online session. We then resume the search from the \textit{RouterInfo} following the last routine data to identify the next session using the same approach. Additionally, any non-routine \textit{RouterInfo} between two sessions is assigned to the preceding session if its time gap from the session's last routine data is shorter than the routine interval.\looseness=-1

Relying solely on routine \textit{RouterInfos} may miss temporary offline behaviors due to random \textit{RouterInfo} update intervals, which can vary by up to 10 minutes. As a result, \textit{RouterInfo} from adjacent online sessions may overlap with routine intervals during temporary offline behavior, causing the coarse-grained method to misidentify session boundaries. This leads to two types of errors: \textbf{mismatch} and \textbf{misidentification} of routine \textit{RouterInfo}. Mismatch errors occur when the interval between the last routine \textit{RouterInfo} from the previous session ($R_p^r$) and a \textit{RouterInfo} from the next session ($R_s$) coincides with the routine interval. This leads to two distinct results, depending on the characteristic of $R_s$: (1) If $R_s$ is non-routine (\textbf{Case 1} in \Cref{tab:special_cases}), $R_s$ is mistakenly assigned to the previous session, causing incorrect inferences about the previous session's end time and the subsequent session's start time. (2) If $R_s$ is routine (\textbf{Case 2} in \Cref{tab:special_cases}), the two sessions are incorrectly merged, missing the offline period in between. Misidentification occurs when a non-routine \textit{RouterInfo} after the last routine \textit{RouterInfo} in the previous session ($R_p$) and $R_s$ from the next session are mistakenly treated as successive routine publications. Since a floodfill router always sends a non-routine \textit{RouterInfo} before it goes offline (we discuss it in our leave behavior identification method), this error generally occurs when dealing with floodfill routers, leading to two possible outcomes: (1) If $R_s$ is also non-routine (\textbf{Case 3} in \Cref{tab:special_cases}), $R_p$ and $R_s$ are incorrectly treated as part of the same session, splitting two sessions into three. (2) If $R_s$ is routine (\textbf{Case 4} in \Cref{tab:special_cases}), $R_p$ is assigned to the next session, also causing incorrect inferences about the previous session's end and the subsequent session's start times. To address these challenges, we propose a fine-grained join behavior identification method for online sessions. \looseness=-1

\textbf{Join behavior identification for Java-based routers.} To enable fine-grained online session inference, we analyze the publication pattern of Java-based routers to identify their join behavior during startup. The routine \textit{RouterInfo} is published every 4 update intervals, with a state variable $C$ counting the tasks since startup. Specifically, \textit{RouterInfo} is published when $C~ \text{mod}~4 = 0$. Upon startup, $C$ starts at 0, and the first \textit{RouterInfo} is published within 10 seconds, incrementing $C$ to 1. The next task is scheduled 90 seconds later but cannot execute until 9 minutes after the latest upload. If the first upload succeeds within 90 seconds (i.e., the acknowledgment message for \textit{RouterInfo} is received), the second task executes between 9 and 10.5 minutes, resulting in an interval of $3 \times D_c$ (25-31.5 minutes) between the first and first routine \textit{RouterInfo}. Alternatively, if the router fails to receive the acknowledgment within 90 seconds, $C$ increments to 2 after approximately 90 seconds. After two more update intervals, $C$ reaches 4, triggering the first routine \textit{RouterInfo}, with the interval between the first and first routine \textit{RouterInfo} being approximately $90 + 2 \times D_c$ (17.5-22.5 minutes). \looseness=-1

\begin{figure}[t]
    \centering
    \includegraphics[width=0.8\linewidth]{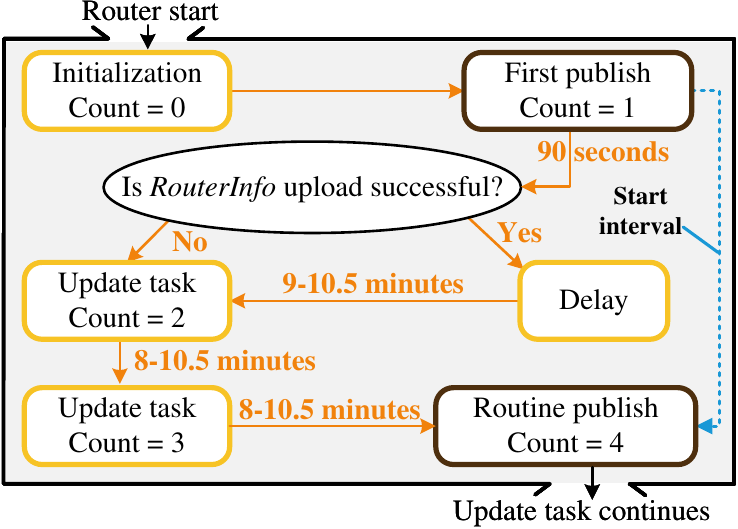}
    \caption{Startup workflow of Java-based I2P routers.}
    \label{fig:startup_publish}
\vspace{-5mm}
\end{figure}

By analyzing the publication pattern during the startup phase of Java-based routers, we can accurately determine the start time of each online session. Starting from the first routine \textit{RouterInfo} of an online session, we scan the entire trace for \textit{RouterInfo} entries that match one of two possible join behaviors (i.e., a time interval of $3 \times D_c$ or $90 + 2 \times D_c$) to identify the first \textit{RouterInfo} published after the router's startup. This approach allows for fine-grained identification of join behaviors, effectively addressing challenges posed by temporary offline behaviors. \looseness=-1

While the join behavior identification method effectively corrects all the issues introduced by the coarse-grained method when dealing with the router's temporary offline behavior, it requires a complete join behavior---characterized by a distinct interval between two \textit{RouterInfo} publications. If the router briefly returns online and publishes only one \textit{RouterInfo}, the join behavior is incomplete (\textbf{Case 5} in \Cref{tab:special_cases}), rendering the method inapplicable. In this case, we instead rely on the leave behavior identification method to correctly infer the end of the previous online session, and thereby infer the correct start time of the subsequent online session. \looseness=-1

\textbf{Leave behavior identification for Java-based routers.} We find that only Java-based floodfill routers display explicit leave behavior. Immediately before shutdown, they publish a final \textit{RouterInfo} whose \textit{options} field removes the `f' flag, indicating they will cease floodfill service. By scanning backward from the last routine \textit{RouterInfo} in an online session, we locate this publication and use its timestamp as the router's leave time. This allows us to identify a router’s offline behavior even if its join behavior is lost or absent in subsequent sessions. For non-floodfill routers, which lack explicit offline patterns, we estimate offline times to reduce discrepancies, as discussed in \S\ref{subsec:complement}.\looseness=-1

\subsection{Online Session Complement}
\label{subsec:complement}

While our live behavior inference method can theoretically group online sessions accurately, the low-cost deployment of floodfill routers leads to incomplete \textit{RouterInfo} collection in practice. To address this issue, we verify the integrity of each identified session, detect missing join or leave behaviors, and supplement the missing information by analyzing the publication context. Although the analytical principles are consistent for both Java- and C++-based routers, the events that cause missing \textit{RouterInfo} differ across implementations. Therefore, to avoid redundancy, we present the Java-based techniques in the main text and provide the corresponding C++-specific procedures in Appendix \ref{subsec:C++-based_complement}. \looseness=-1

\begin{figure}[t!]
    \centering
    \includegraphics[width=0.85\linewidth]{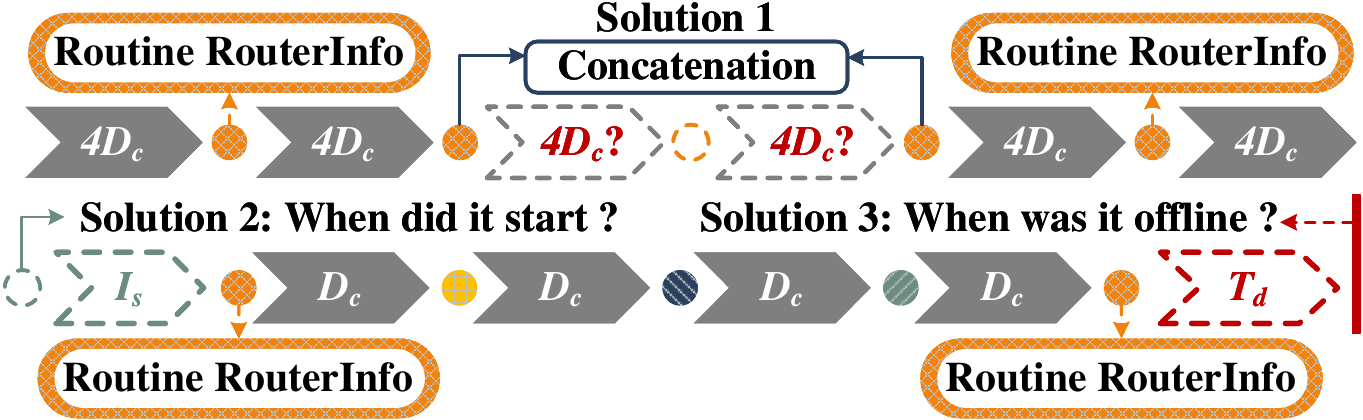}
    \caption{Three solutions for online session complement.}
    \label{fig:deviation_control}
\vspace{-5mm}
\end{figure}

Missing \textit{RouterInfo} can lead to three issues: loss of routine \textit{RouterInfo} (\textbf{Cases 6--8 and 14} in \Cref{tab:special_cases}), join behavior (\textbf{Cases 10--12 and 15}), and leave behavior (\textbf{Cases 9 and 16}), each causing errors in session inference. 
In the first issue, missing consecutive routine \textit{RouterInfo} splits a session, leaving the second session without join behavior. In the second and third issues, missing data indicates the absence of join or leave behavior. We propose three solutions to address each issue and present how to handle online session inference errors in these situations, as shown in \Cref{fig:deviation_control}.\looseness=-1

\textbf{Solution 1: Online session concatenation.} 
When join behavior is missing in both the first and second issues, we attempt to concatenate consecutive sessions to verify whether they belong to the same one (\textbf{Case 8} in \Cref{tab:special_cases}). For Java-based routers, if the time interval between the final routine \textit{RouterInfo} of the preceding session and the initial routine \textit{RouterInfo} of the succeeding session matches a multiple of the routine interval, the sessions are merged. This condition can be expressed as:
\vspace{-1mm}
\begin{equation}
    \label{eq:concatenate_restriction}
    \begin{aligned}
        \left\{
            \begin{array}[pos]{ll}
                \min{(D_c)} \leqslant \frac{I_k}{m_t} \leqslant \max{(D_c)} &           \\
                m_t = 4 \times k,                                                            & k = 2,3
            \end{array}
        \right. 
    \end{aligned}
\end{equation}
where $I_k$ is the interval between two routine \textit{RouterInfo}, and $m_t$ is the number of update intervals within $I_k$. We limit $I_k$ to a maximum of three routine intervals, as empirical testing shows no instance of consecutively losing more than two routine \textit{RouterInfo}. \looseness=-1

\textbf{Solution 2: Join behavior supplementation.} If a session has no join behavior and cannot be merged using \textbf{Solution 1}, the join behavior is considered missing or incomplete. For Java-based routers, the join behavior consists of two \textit{RouterInfos}. When one is missing, we determine which publication is absent and infer its expected position based on the characteristic startup interval. 
The initial \textit{RouterInfo} should be published an interval before routine behavior begins, referred to as the \textit{startup interval}, measured as either $3 \times D_c$ or $90 + 2 \times D_c$, denoted by $D_{s1}$ and $D_{s2}$, respectively. If the first routine \textit{RouterInfo} is missing (\textbf{Case 6} in \Cref{tab:special_cases}), the expected gap between the initial and the next routine publication becomes the sum of the startup and routine intervals---either $7 \times D_c$ or $90 + 6 \times D_c$. We thus search backward from the first captured routine \textit{RouterInfo} within this time window to locate a potential initial \textit{RouterInfo}. If found, its publication time is regarded as the session's start time. If not, we infer that the missing data is the initial \textit{RouterInfo} (\textbf{Case 10} in \Cref{tab:special_cases}).
In this case, we compute the expected values for $D_{s1}$ and $D_{s2}$ as follows:
\begin{equation}
    \label{eq:start_expect_cases}
    \begin{aligned}
        \renewcommand{\arraystretch}{1.2}
        \left\{
            \begin{array}[pos]{l}
                E(D_{s1}) = 90 + 2\times E\left(D_c\right) \\
                E(D_{s2}) = 3\times E\left(D_c\right) \\
            \end{array}
        \right. 
    \end{aligned}
\end{equation}
where $E\left(D_c\right)$ is the expected duration of an update interval, given by $\frac{3T+4E(S)}{16}$. According to \Cref{eq:check_interval}, and given that the probabilities of these two cases are generally equal, the expected value of the startup interval, $E_f$, is calculated as
\begin{equation}
    \label{eq:start_expectation}
    E_f = \frac{E(D_{s1})}{2} + \frac{E(D_{s2})}{2} = 45 + \frac{15T+20E(S)}{32}
\end{equation}
Thus, the start time of the online session can be estimated as $E_f$ earlier than the publication time of the first routine \textit{RouterInfo}. \looseness=-1

\textbf{Solution 3: Leave behavior supplementation.} Our goal is to estimate the expected remaining online duration after the router’s final observed \textit{RouterInfo} (\textbf{Case 9} and \textbf{16} in \Cref{tab:special_cases}). For Java-based routers, non-floodfill routers do not exhibit explicit leave behavior, while floodfill routers may lose the \textit{RouterInfo} without a `f' flag, which serves as an indicator of leave behavior. In both cases, since no additional routine \textit{RouterInfo} is received after the router’s last \textit{RouterInfo}, we can approximate that the router goes offline before the next routine \textit{RouterInfo} would appear. We first estimate the expected publication time for the next routine \textit{RouterInfo} $T_n$ by:
\begin{equation}
    \label{eq:routine_exp}
    T_n=T_r + 4\times E(D_c)
\end{equation}
where $T_r$ denotes the publication time of the last routine \textit{RouterInfo}. Then the inferred offline time $T_o$ is given by 
\begin{equation}
\label{eq:last_duration}
    T_o=\frac{T_n-T_f}{2}+T_f=\frac{T_r+T_f}{2}+2E(D_c)
\end{equation}
where $T_f$ represents the publication time of the last \textit{RouterInfo} in the online session. \looseness=-1

\textbf{Discussion:} We analyze the remaining cases (\textbf{Cases 7, 11, and 12} in \Cref{tab:special_cases}) caused by missing \textit{RouterInfo} that cannot be resolved using the proposed solutions. \textbf{Case 7} occurs when the last routine \textit{RouterInfo} for a non-floodfill router is lost. Without explicit leave behavior, this results in insufficient timing information to infer the router’s leave time, leading to an inference bias of about 35--45 minutes. \textbf{Case 11} arises when the initial \textit{RouterInfo} for a Java-based non-floodfill router is lost after a brief offline period, causing an incomplete join behavior. If the interval between the first routine \textit{RouterInfo} of the subsequent session ($R_f^s$) and the last routine data ($R_{\ell}^p$) of the previous session equals the sum of the startup and routine intervals, our \textbf{Solution 2} may incorrectly infer that a routine \textit{RouterInfo} between them is lost, treating $R_{\ell}^p$ as the initial \textit{RouterInfo} of the current session. Although this error cannot be corrected without both join and leave behaviors, the two sessions remain distinguishable, and both inference biases of the previous session's end time and the subsequent session's start time are within a routine interval (less than 40 minutes). If the interval between $R_f^s$ and $R_{\ell}^p$ coincides with the routine interval, the coarse-grained method may incorrectly merge the sessions (\textbf{Case 12}). Although uncorrectable, this case is rare due to the specific timing requirement. \looseness=-1

\subsection{Live Behavior Probing for Hidden Service}
\label{sec:server_state}

\begin{figure}[t]
    \centering
    \includegraphics[width=0.8\linewidth]{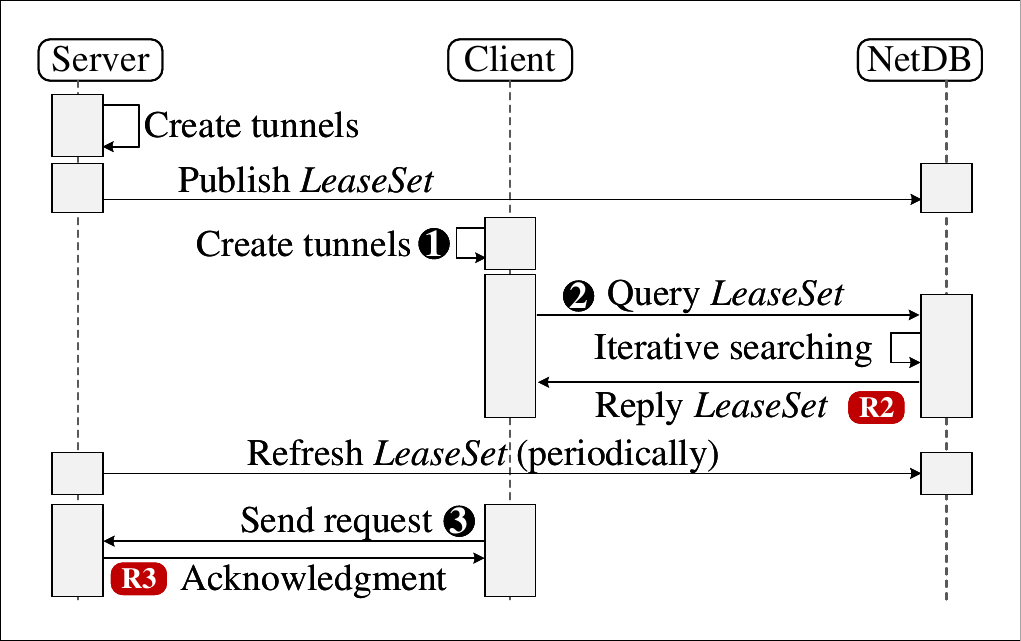}
    \caption[short]{Process of accessing a hidden service.}
    \label{fig:HS_access}
\vspace{-5mm}
\end{figure}

Unlike the passive method for routers, we use an active probing approach to infer the live behavior of a target hidden service. As shown in \Cref{fig:HS_access}, the process for an I2P client to access a hidden service involves three stages: (i) the client establishes a pair of unidirectional tunnels for communication. (ii) The client queries the netDB for the target hidden service’s LeaseSet. If the response message $R2$ includes the \textit{LeaseSet}, the lookup is successful; otherwise, it indicates that the \textit{LeaseSet} is not present in the netDB. (iii) Using the \textit{LeaseSet}, the client connects to the hidden service as depicted in \S\ref{subsec:garlic} and sends a service request, receiving acknowledgment via message $R3$. \looseness=-1

Based on this mechanism, the hidden service's live behavior can be inferred from the messages at each stage. If a failure response $R2$ is received, it indicates the hidden service has been offline for more than 10 minutes, causing the \textit{LeaseSet} to expire. If the \textit{LeaseSet} is retrieved but no connection is made, the service is inferred to have been offline within the last 10 minutes. Successfully receiving $R3$ confirms the service's availability. By periodically repeating this access task, the live behavior of the hidden service can be inferred. Due to the I2P network's inherent anonymity, our client remains unexposed to the hidden service. In addition, I2P reliably delivers probing messages to the target service by using the SSU (Secure Semireliable UDP) and NTCP (NIO-based TCP) transport protocols, as well as the application-layer streaming library. \looseness=-1

The network environment has limited impacts on the attack's effectiveness. I2P provides approaches to ensure the reliability of data transmission. At the transport layer, I2P implements the SSU and NTCP protocols based on UDP and TCP, respectively, to ensure reliable message delivery. At the application layer, I2P provides the streaming library for the same purpose. Because of the above strategies, as long as the host router’s RouterInfo and the hidden service’s response to probing by the attacker could be delivered correctly, our attack can succeed.

\subsection{Live Behavior Correlation}
\label{subsec:similar_alg}

We propose a dynamic time warping (DTW)-based correlation method to evaluate the similarity between the live behaviors of routers and a target hidden service. First, we serialize the inferred live behavior by representing online and offline states with positive and negative integers, respectively, while recording the duration of each state. For example, if a router stays online for 5 minutes and then offline for 5 minutes, its state sequence would be: {1, 2, 3, 4, 5, -1, -2, -3, -4, -5}. The DTW algorithm then computes the alignment cost between two sequences by calculating the pairwise distance between corresponding elements. Unlike traditional methods that rely on predefined distance metrics (e.g., Euclidean distance), we use a customized distance function: if two numbers are identical, the distance is 0; if they share the same sign, the distance is 1; and if they have opposite signs, the distance is 2. The DTW algorithm outputs a distance score, where a lower score indicates higher similarity between the two live behaviors.\looseness=-1

\subsection{Theoretical Analysis}
\label{subsec:theoretic}

We now prove that a target hidden service exhibits a unique live behavior pattern over time and can therefore be uniquely identified if this behavior is measurable, as demonstrated in this paper.
We can model the live behavior of a hidden service as an on-off stochastic process. Therefore, $n$ hidden services correspond to $n$ on-off stochastic processes. Assume that their $on$ periods follow the same discrete probability distribution and have $k$ values with probabilities $a_1, \dots, a_k$. Their $off$ periods follow the same discrete probability distribution and have $l$ values with probabilities $b_1, \dots, b_l$. While one might assume the $n$ hidden services are more distinct, this model assumes they are more similar, presenting a worse-case scenario for the attacker.
Assume that the $n$ processes are independent and their on-off cycles are independent, where one cycle contains one $on$ period and one $off$ period. \looseness=-1

\begin{theorem}
\label{thm::PA}
The probability $P_{\mathcal{A}}$ that one particular on-off process $A$ does not produce the same length-m on-off sequence with any of the other on-off $n-1$ processes can be calculated as follows,
\begin{equation}
\label{eqn::PA}
P_{\mathcal{A}}= \big( 1 - (S_a S_b)^m \big)^{n-1},
\end{equation}
where 
\begin{equation}
S_a = \sum_{i=1}^k a_i^2, \qquad
S_b = \sum_{j=1}^l b_j^2.
\end{equation}
\end{theorem}
The detailed derivation can be found in \Cref{sec:ProbabilityAnalysis}.
It can be observed that in \Cref{eqn::PA}, $S_a S_b <1$. Therefore, when $m$ is large, $P_{\mathcal{A}}$ approaches 1. That is, we can uniquely identify a hidden service given a long observation of its live behavior. \looseness=-1

\begin{corollary}
\label{corollary::SynchronizedProcesses}
If $\mathcal{A}$ and the other on–off processes are synchronized in their on–off behavior, they become identical and cannot be distinguished based solely on that behavior.
\end{corollary}

\section{Evaluation}
\label{sec:eval}
We implemented \sys and conducted extensive experiments within the real-world I2P network under controlled conditions. In this section, we first present details of our experimental setup and then present the experimental results, addressing three research questions as follows:

\begin{itemize} [label={},left=0pt, leftmargin=1em, itemindent=-1em]

    \item \textbf{(RQ1):} Are the inferred live behaviors of I2P routers consistent with their actual behaviors using \textit{RouterInfo} data (\S\ref{subsubsec:exp_consistency})?
    
    \item \textbf{(RQ2):} Can our method accurately reconstruct the router's live behavior when we have incomplete data (\S\ref{subsubsec:exp_reliability})?
    
    \item \textbf{(RQ3):} Can \sys effectively deanonymize the target hidden service (\S\ref{subsubsec:Effectiveness})?

\end{itemize}

\subsection{Experiment Setup}
\label{subsec:exp_setup}

\textbf{Experiment environment.} We deploy 25 I2P routers on virtual private servers (VPS) running Ubuntu 22.04, distributed across various regions. Each VPS is configured with 1 virtual CPU, 1 GB RAM, and 32 GB storage. All routers use customized I2P software to collect \textit{RouterInfo} data and log essential information without disrupting normal network operations. 15 routers operate in floodfill mode to support large-scale \textit{RouterInfo} collection, with the sufficiency of 15 floodfill routers demonstrated in \Cref{subsec:floodfill_num}. The remaining 10 routers host the hidden services used as targets in our deanonymization experiments: 5 Java-based and 5 C++-based. Among the Java-based routers, 3 have reachable ports and two are firewalled. One of the reachable Java routers is configured in the floodfill mode to validate our leave behavior identification method. All five C++-based routers are configured in the non-floodfill mode, with 3 having reachable ports and 2 firewalled. An I2P client is deployed on a Windows 11 machine (Intel Core i7-1165G7, 16 GB RAM) to probe the controlled hidden services and process data for the deanonymization prototype. \looseness=-1

\begin{table}[!b]
\centering
\scriptsize{
    \caption{Live behavior settings for different scenarios.}
    \setlength{\tabcolsep}{1.2pt}
    \begin{tabular}{M{1cm}| L{7.6 cm}}
        \toprule[1.5pt]
        \multicolumn{2}{c}{\textbf{Hidden services with frequent on-off behaviors (Time values are in minutes)}} \\
        \midrule
        \textbf{S1} & 100, -45, 110, -50, 120, -55, 130, -60, 100, -55, 110, -50, 120, -45, 130, -160 \\
        \textbf{S2} & 120, -30, 110, -35, 100, -40, 120, -45, 110, -40, 100, -30, 10, -35, 120, -30, 10, -35, 100, -40, 120, -60 \\
        \textbf{S3} & 140, -10, 130, -5, 120, -10, 110, -5, 10, -10, 100, -10, 130, -5, 10, -10, 120, -10, 110, -5, 100, -10, 110, -5, 10, -10, 120, -15 \\
        \textbf{S4} & 160, -20, 160, -15, 160, -10, 10, -5, 160, -30, 160, -40, 10, -10, 160, -50, 160, -60, 10, -50 \\
        \midrule
        \multicolumn{2}{c}{\textbf{Hidden services with stable live behaviors}} \\        
        \midrule
        \textbf{S5} &  Online for nine days; Offline on Day 10 \\        
        \textbf{S6} &  Online for 28 days; Offline on Day 29 \\        
        \textbf{S7} & Online for 50 days; Offline on Day 51 \\        
        \bottomrule[1.5pt]
    \end{tabular}
\label{tab:behavior_setting}
}
\end{table}

\begin{figure*}[!h]
    \centering

    \subfloat[]{\includegraphics[width=0.245\textwidth]{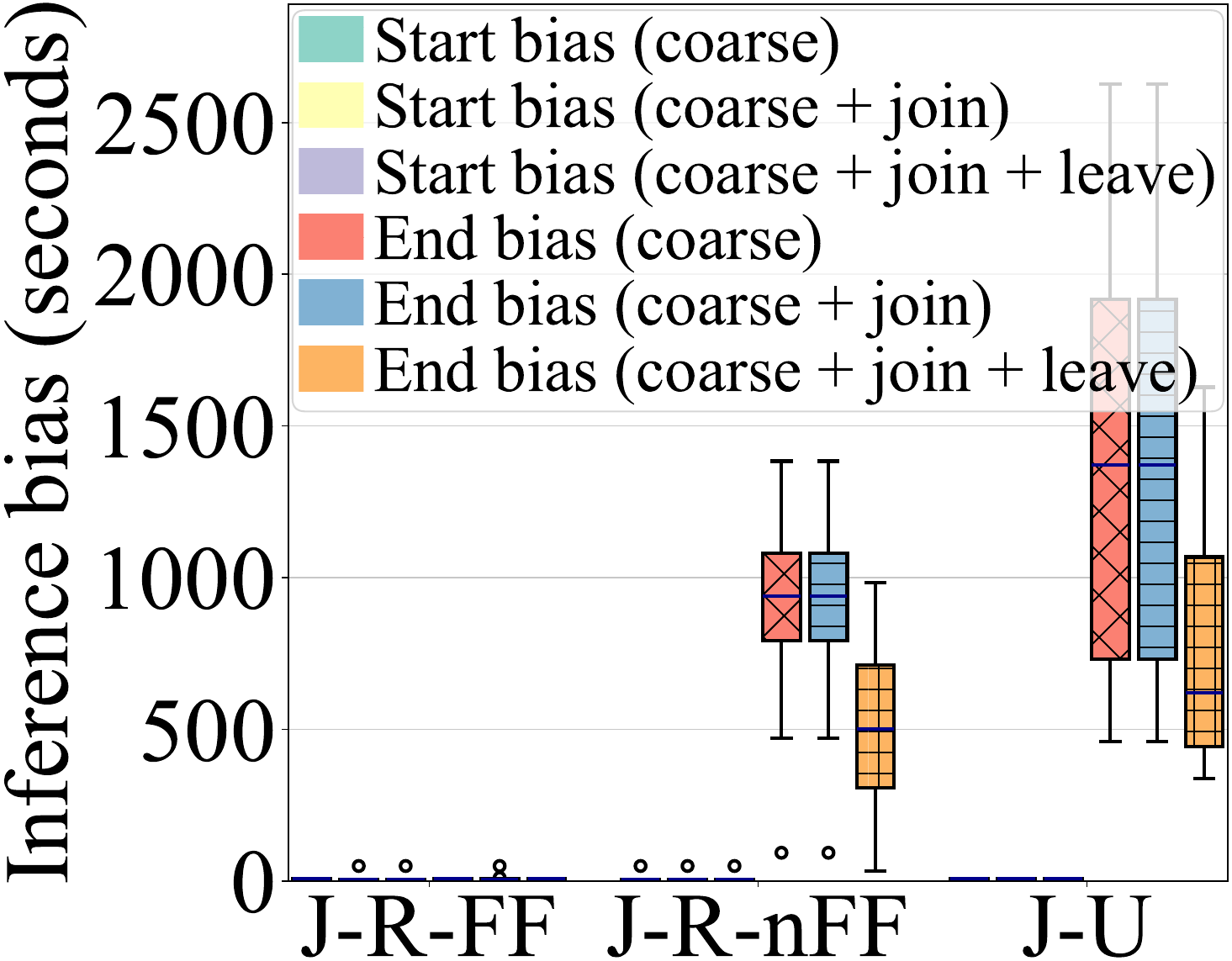}%
    \label{fig:45-60_result}}
    \hfill
    \subfloat[]{\includegraphics[width=0.245\textwidth]{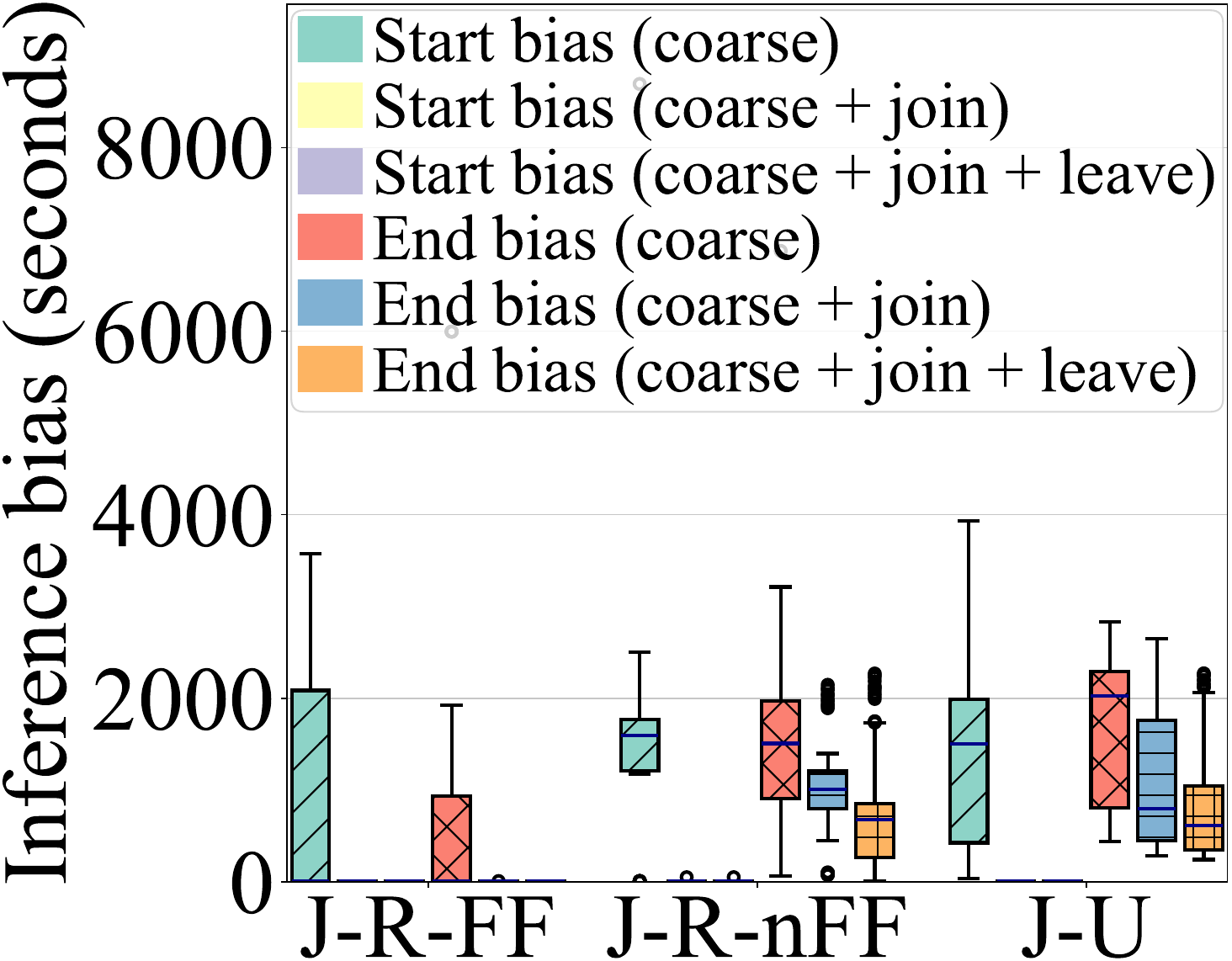}%
    \label{fig:30-45_result}}
    \hfill
    \subfloat[]{\includegraphics[width=0.245\textwidth]{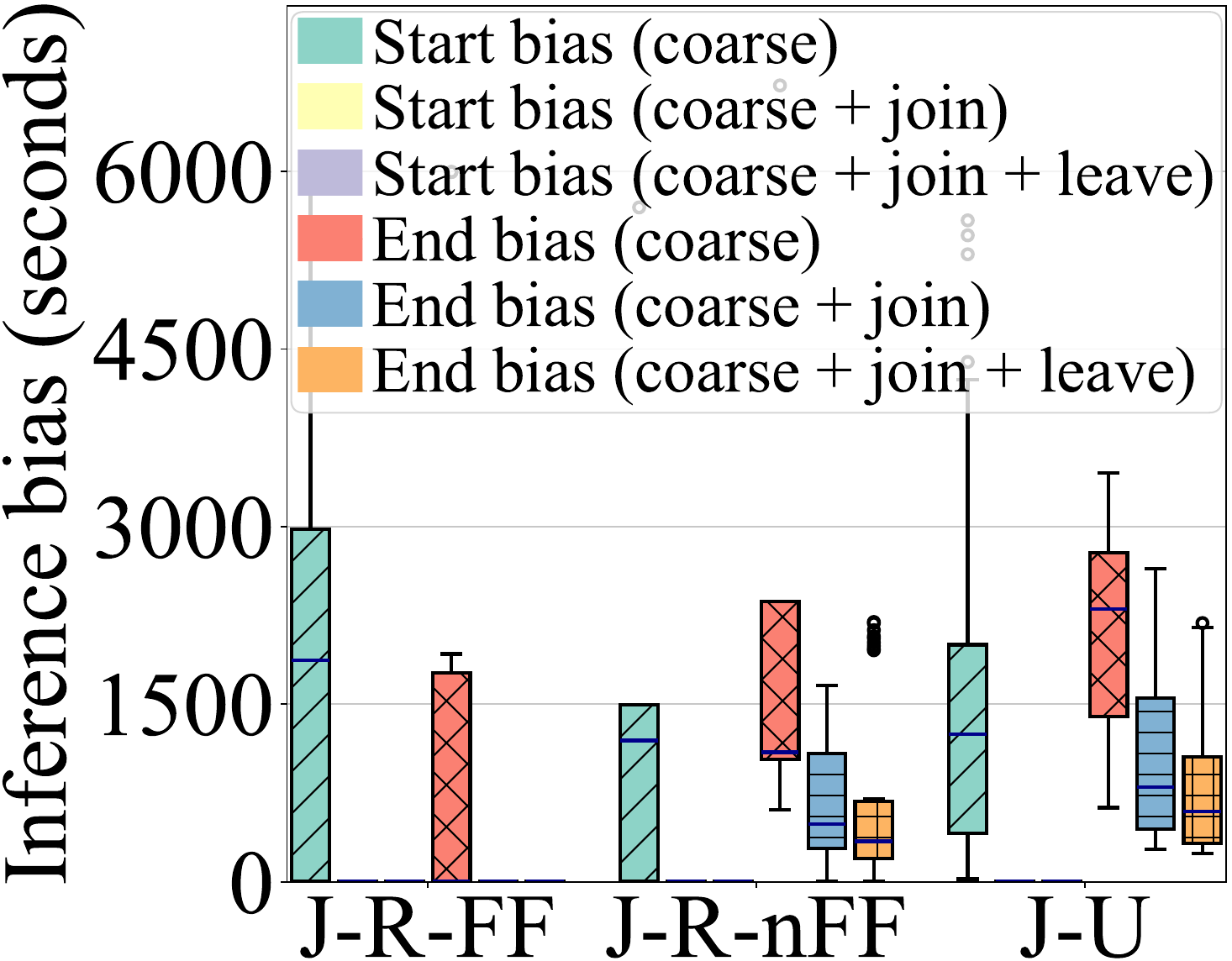}%
    \label{fig:5-10_result}}
    \hfill
    \subfloat[]{\includegraphics[width=0.245\textwidth]{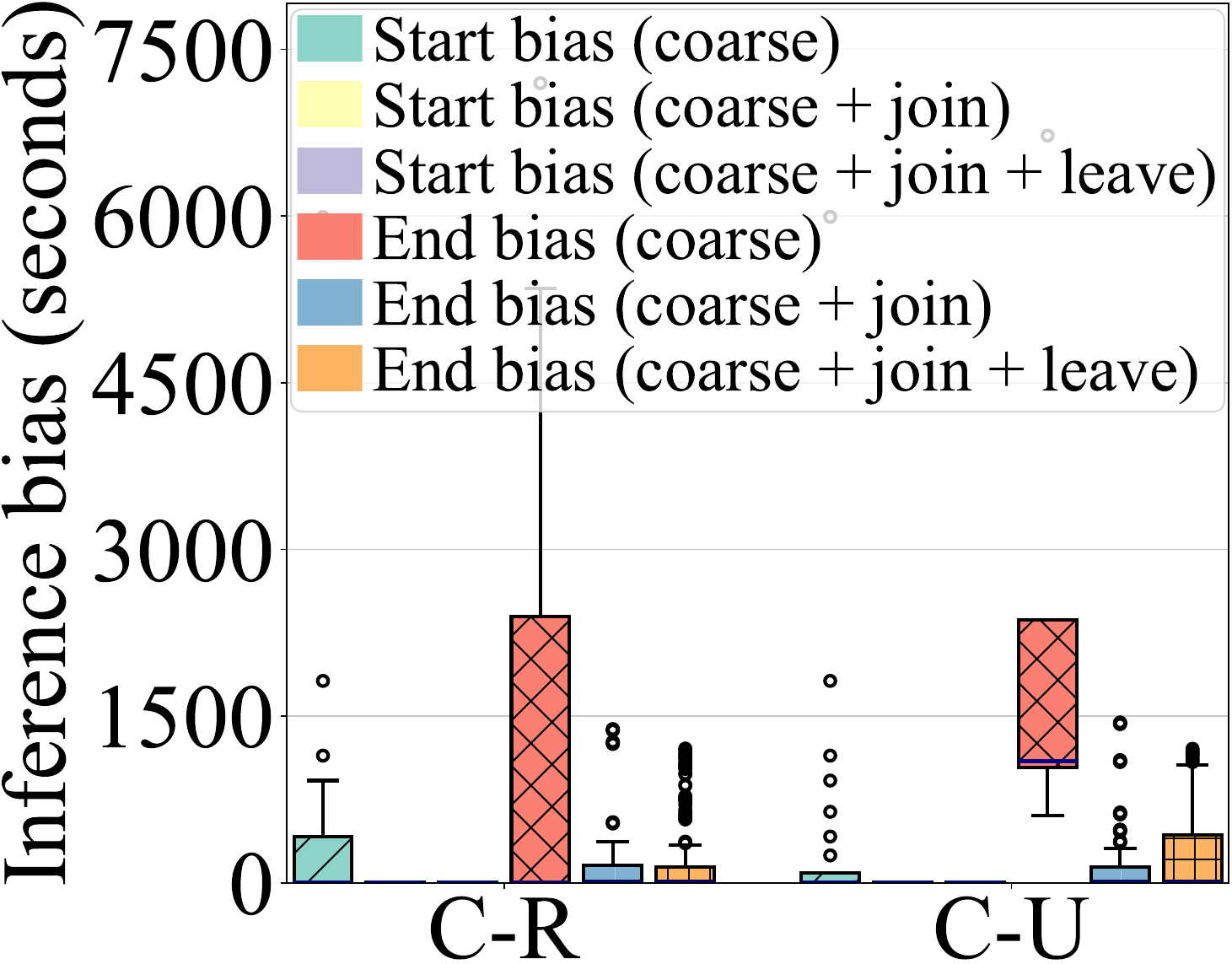}%
    \label{fig:C_result}}

    \caption{Results of the ablation experiments. Subfigures (a–c) show results for Java-based routers under different offline durations: (a) 45–60 min (S1), (b) 30–45 min (S2), and (c) 5–10 min (S3). Subfigure (d) presents results for C++-based routers with mixed offline durations (S4). J-R-FF, J-R-nFF, and J-U denote Java-based floodfill, non-floodfill, and unreachable-port routers, respectively. C-R and C-U refer to C++-based routers with reachable and unreachable ports.\looseness=-1}
    \label{fig:complete_data_results}
    \vspace{-4mm}
\end{figure*}

\textbf{Dataset Collection.} We collect \textit{RouterInfo} data from 15 controlled floodfill routers over a period of more than 8 months. To create a \textit{ground truth dataset}, we locally record the \textit{RouterInfo} publication logs from 10 host routers, resulting in a complete, lossless dataset. This ground truth dataset serves to evaluate the contribution of each component in the fine-grained inference method and to validate the accuracy of inferred live behavior under ideal conditions. Additionally, we capture \textit{RouterInfos} from the 15 deployed floodfill routers to build an \textit{evaluation dataset}, which may experience partial data loss due to network dynamics. This dataset is used to assess both the necessity of the online session complement strategy and the overall effectiveness of the live behavior inference framework under realistic conditions. These two datasets allow us to evaluate the performance of our inference method in both ideal (no data loss) and realistic (with data loss) settings. \looseness=-1

\textbf{Controlled Online/Offline Behavior Scenarios.} 
We configure the online/offline behaviors of the controlled host routers and consider seven scenarios as shown in \Cref{tab:behavior_setting} to evaluate the accuracy of our live behavior inference method.
Router behaviors are controlled using \textit{crontab} scripts to follow the specified configurations automatically.
\textbf{S1} to \textbf{S4} represent hidden services that exhibit frequent on–off patterns, posing challenges for inferring the live behavior of their host routers, specifically, accurately mapping each \textit{RouterInfo} to its associated online session. Services in \textbf{S1} to \textbf{S4} run for two months. Their daily behavioral patterns are defined in \Cref{tab:behavior_setting}. Positive values indicate the duration of online sessions, while negative values represent offline periods, with their absolute values denoting the length.
\textbf{S5} to \textbf{S7} correspond to long-lived hidden services, with the hidden services in \textbf{S5} to \textbf{S7} run for the number of days specified in \Cref{tab:behavior_setting}.

For scenarios of hidden services with frequent on-off behaviors, \textbf{S1} to \textbf{S3} target the services hosted on Java-based routers. \textbf{S1} tests the coarse-grained method's ability to segment online sessions, with routers remaining offline for 45–60 minutes, longer than the routine publication interval but shorter than the \textit{RouterInfo} expiration, causing a false online identification by floodfill routers. \textbf{S2} introduces two types of errors (routine \textit{RouterInfo} mismatches and misidentifications) to evaluate the effectiveness of the join behavior identification method, with offline durations set to 30–45 minutes, aligning with the routine publication interval. \textbf{S3} evaluates the leave behavior identification method by introducing incomplete join behaviors, with offline durations reduced to 5–10 minutes, increasing the likelihood of mistakenly merging sessions due to a short offline period. 
\textbf{S4} focuses on the services hosted on C++-based routers, where the offline duration has a minimal impact on the coarse-grained method. This is due to the lack of randomness in routine publication intervals, which eliminates the mismatch and misidentification issues commonly encountered with Java-based routers. Thus, we randomly select offline durations between 5 and 60 minutes to create a varied environment for evaluating the inference method. \looseness=-1

For scenarios of long-lived hidden services, in \textbf{S5}, a hidden service stays online for nine days and goes offline on Day 10. In \textbf{S6}, a hidden service remains online for 28 days before going offline on Day 29. In \textbf{S7}, a hidden service runs continuously for 50 days before shutting down on Day 51. \looseness=-1

\subsection{Experiment Results}
\label{subsec:exp_results}

\subsubsection{Consistency (\textit{RQ1})}
\label{subsubsec:exp_consistency}

\begin{table}[!h]
\centering
\scriptsize{
    \caption{Occurrence of each case in \Cref{tab:special_cases}, triggered by different types of routers under various scenarios. In S1-S3, R-FF, R-nFF, and U denote floodfill, non-floodfill, and unreachable-port routers, respectively. In S4, R and U refer to C++-based routers with reachable and unreachable ports.}

    \setlength{\tabcolsep}{1.2pt}
    \begin{tabular}{l R{0.6cm} R{0.65cm} R{0.5cm} R{0.6cm} R{0.65cm} R{0.5cm} R{0.6cm} R{0.65cm} R{0.5cm} R{0.6cm} R{0.6cm}}
        \toprule[1.5pt]
         \multirow{2}{*}{} & \multicolumn{3}{c}{\textbf{S1: 45-60 (J)}} & \multicolumn{3}{c}{\textbf{S2: 30-45 (J)}} & \multicolumn{3}{c}{\textbf{S3: 5-10 (J)}} & \multicolumn{2}{c}{\textbf{S4: MIX (C)}}\\
        \cmidrule(lr){2-4} \cmidrule(lr){5-7} \cmidrule(lr){8-10} \cmidrule(lr){11-12}
         & R-FF & R-nFF & U & R-FF & R-nFF & U & R-FF & R-nFF & U & R & U\\
        \midrule
        Case 1  & - & - & - & 77  & 152 & 228 & 76  & 86  & 256 & -  & - \\
        
        Case 2  & - & - & - & 19  & 22  & 41  & 117 & 137 & 178 & -  & - \\
        
        Case 3  & - & - & - & 252 & -   & -   & -   & -   & -   & -  & - \\
        
        Case 4  & - & - & - & -   & -   & -   & 161 & -   & -   & -  & - \\

        Case 5  & - & - & - & 25  & 23  & 42  & 67  & 67  & 68  & -  & - \\

        Case 13 & - & - & - & -   & -   & -   & -   & -   & -   & 51 & 65\\
        
        \bottomrule[1.5pt]
    \end{tabular}
\label{tab:case_count_ideal}
}
\vspace{-3mm}
\end{table}

\begin{figure*}[!th]
    \centering

    \subfloat[]{\includegraphics[width=0.243\textwidth]{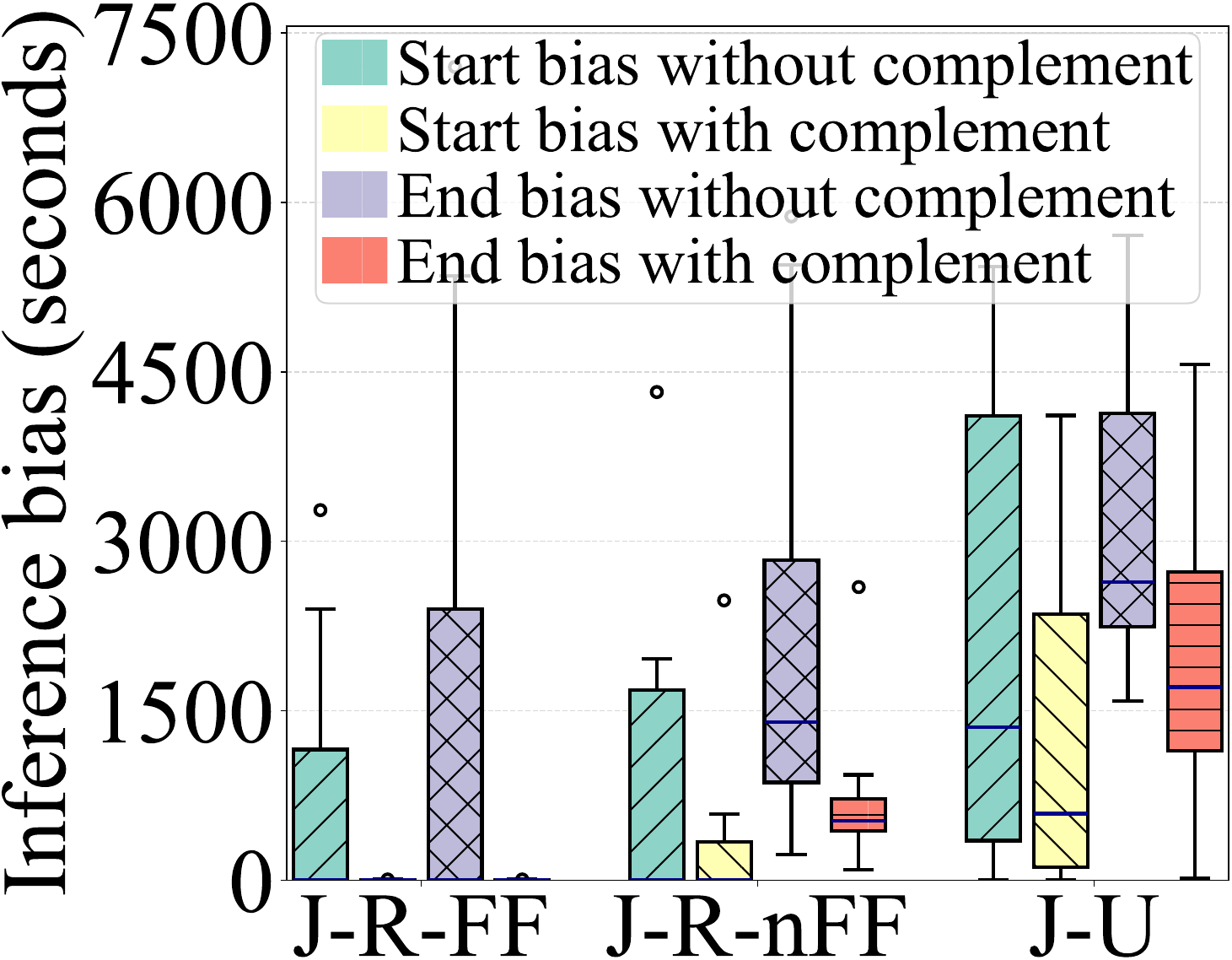}
    \label{fig:45-60_recovery}}
    \hfill
    \subfloat[]{\includegraphics[width=0.243\textwidth]{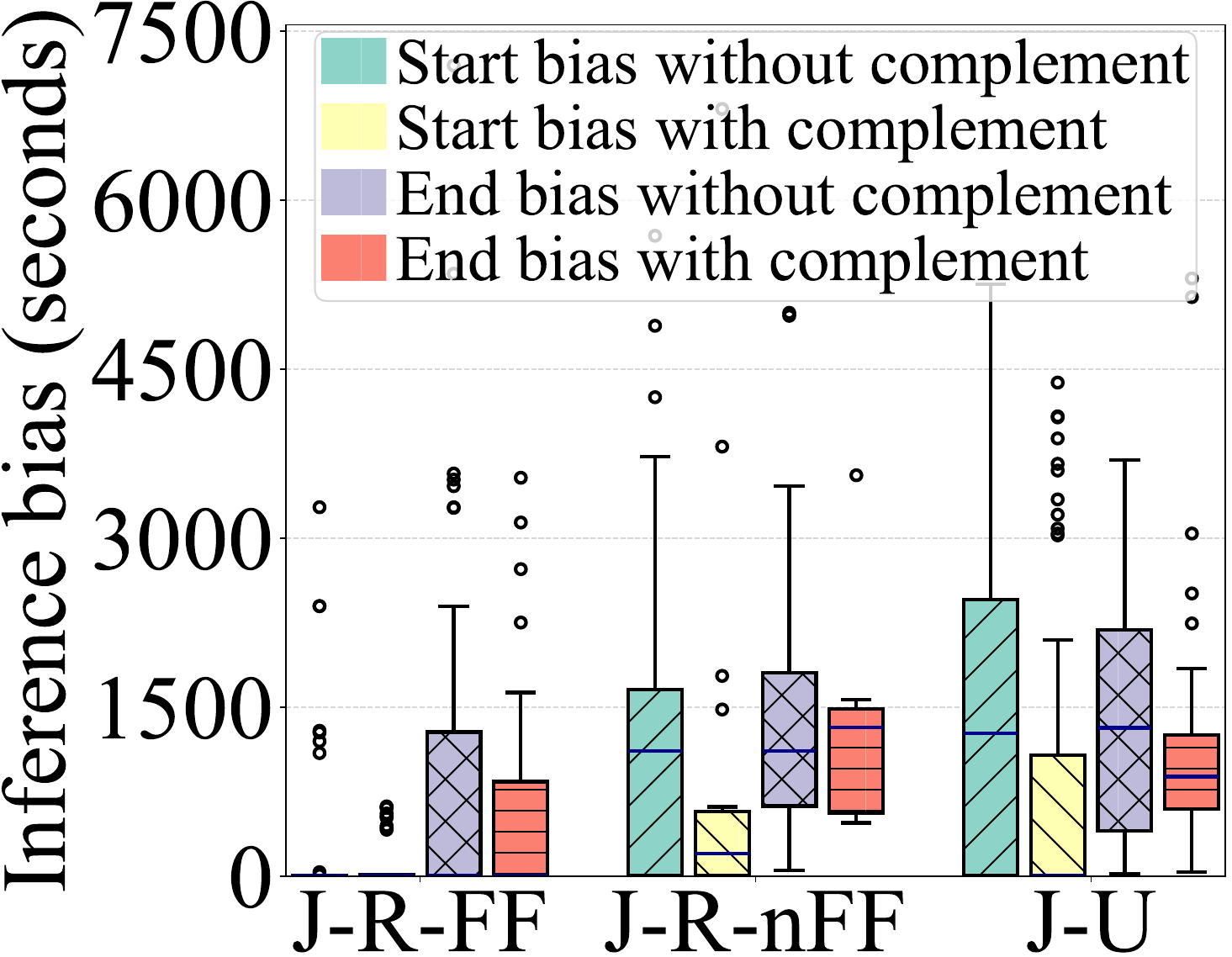}
    \label{fig:30-45_recovery}}
    \hfill
    \subfloat[]{\includegraphics[width=0.243\textwidth]{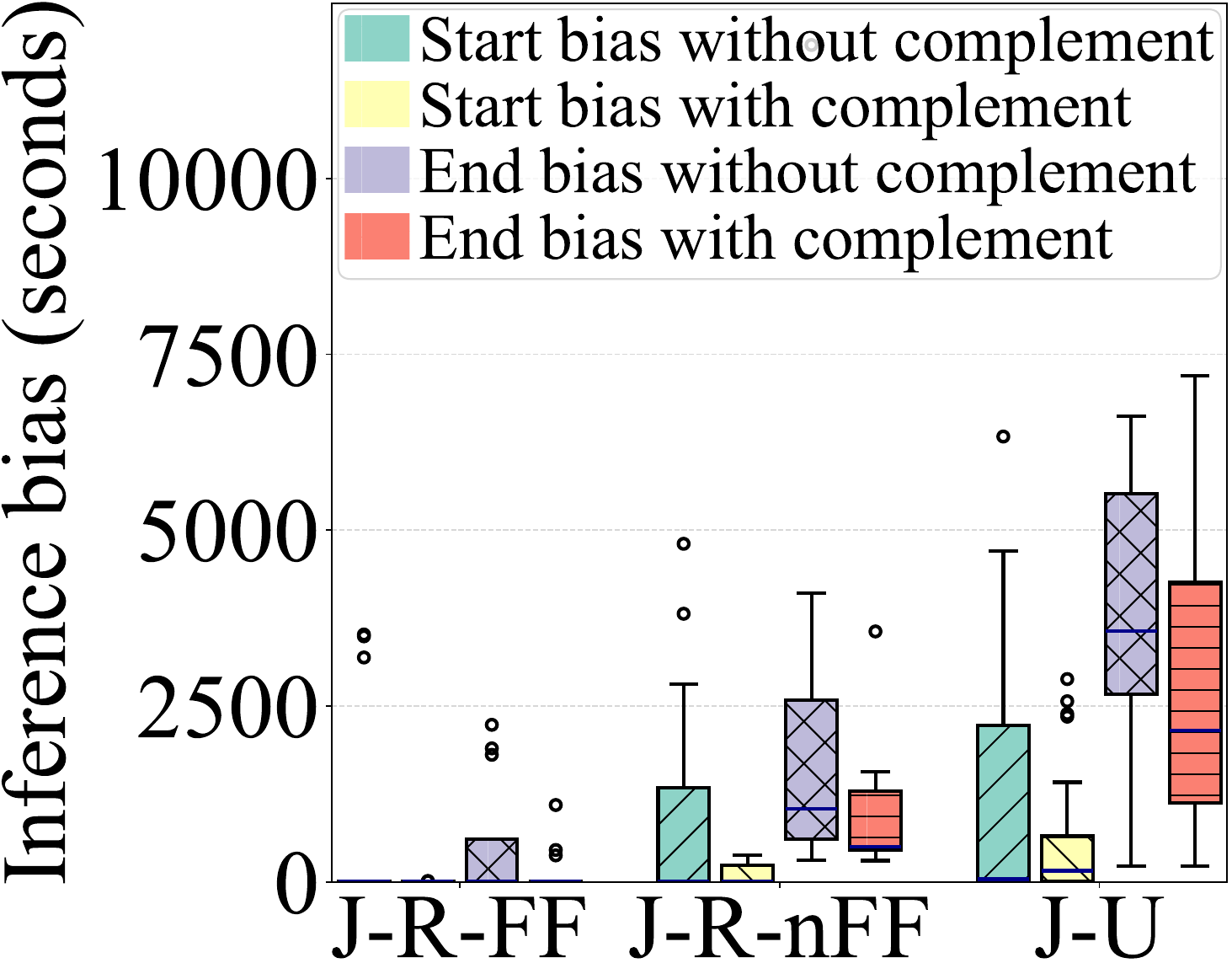}
    \label{fig:5-10_recovery}}
    \hfill
    \subfloat[]{\includegraphics[width=0.243\textwidth]{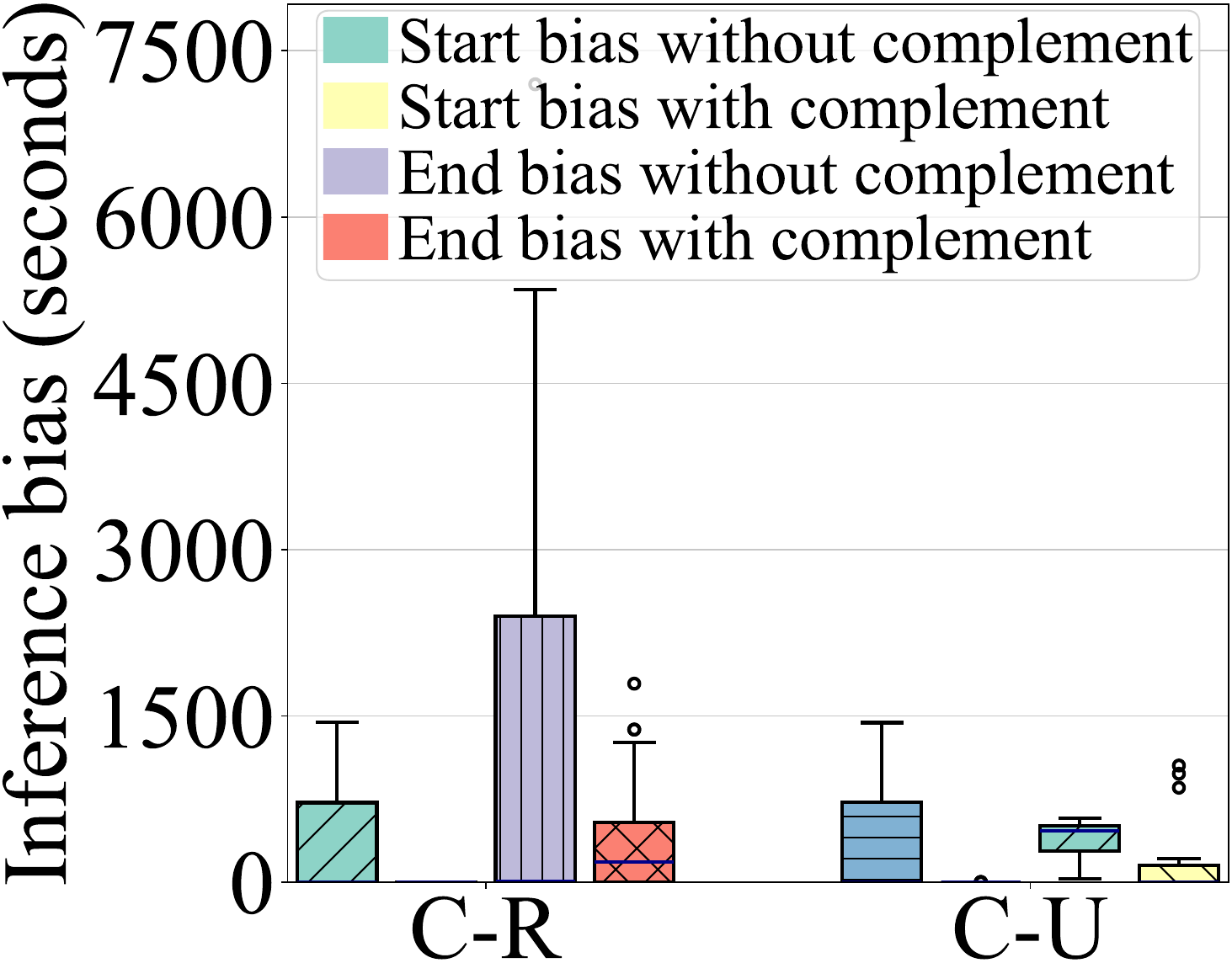}
    \label{fig:C_recovery}}

    \caption{Comparison of live behavior inference biases with and without the online session complement method, using the \textit{evaluation dataset}. Subfigures (a–d) show results under four offline duration settings: (a) S1 (45–60 min), (b) S2 (30–45 min), (c) S3 (5–10 min), and (d) S4 (mixed). Router categories and scenario definitions follow those in \Cref{fig:complete_data_results}.}
    \label{fig:incomplete_data_results}
    \vspace{-4mm}
\end{figure*}

In the experimental scenarios we constructed, along with the corresponding live behaviors assigned to routers, all cases challenging the effectiveness and reliability of the coarse-grained method occur. \Cref{tab:case_count_ideal} reports the occurrences of routine \textit{RouterInfo} mismatches (Cases 1 and 2), misidentifications (Cases 3 and 4), and incomplete join behaviors (Cases 5 and 13), underscoring the need for our proposed join and leave behavior identification methods.

We conduct ablation experiments to evaluate the necessity of each component in our I2P router live behavior inference framework. Beginning with the coarse-grained method, we incrementally incorporate join and leave behavior identification, as well as the leave behavior complement method, to achieve fine-grained inference. Accuracy is measured by the absolute time difference (in seconds) between actual and inferred join and leave times. Results for all four controlled scenarios of \textbf{S1}--\textbf{S4} are presented in \Cref{fig:complete_data_results}. \looseness=-1

\Cref{fig:45-60_result} shows that in \textbf{S1}, the coarse-grained method accurately infers all online session start times, with a median bias of just 6 seconds. However, the accuracy of leave time inference varies across router settings. Since all \textit{RouterInfo} are correctly assigned to their respective sessions and the first \textit{RouterInfo} of each session is accurately identified, the join behavior identification method offers no further improvement. For floodfill routers, the coarse-grained method achieves high accuracy in leave time inference, with a median bias below 10 seconds, as it captures all \textit{RouterInfo} published after the last routine \textit{RouterInfo} within the same session. In contrast, for non-floodfill routers lacking explicit leave behavior, the median leave time bias reaches approximately 900 seconds, which can be reduced to around 500 seconds using the leave behavior complement method.

\Cref{fig:30-45_result} and \Cref{fig:5-10_result} show that the coarse-grained method exhibits significant biases in inferring the join times of controlled routers in scenarios \textbf{S2} and \textbf{S3}, with several exceeding 2000 seconds, despite a median below 10 seconds. After incorporating the join behavior identification method, these biases are dramatically reduced to less than 10 seconds. Likewise, applying the leave behavior complement method to non-floodfill routers cuts the leave time bias by more than half, lowering the median from 1400 seconds to below 700 seconds (12 minutes), which has minimal impact on the similarity between inferred and actual live behaviors. These results confirm the fine-grained method’s effectiveness in enhancing inference accuracy.

\begin{table}[!b]
\vspace{-3mm}
\centering
\scriptsize{
    \caption{Occurrence of each case in \Cref{tab:special_cases}, triggered by different types of routers under various scenarios.}

    \setlength{\tabcolsep}{1.2pt}
    \begin{tabular}{l R{0.6cm} R{0.65cm} R{0.5cm} R{0.6cm} R{0.65cm} R{0.5cm} R{0.6cm} R{0.65cm} R{0.5cm} R{0.6cm} R{0.6cm}}
        \toprule[1.5pt]
         \multirow{2}{*}{} & \multicolumn{3}{c}{\textbf{S1: 45-60 (J)}} & \multicolumn{3}{c}{\textbf{S2: 30-45 (J)}} & \multicolumn{3}{c}{\textbf{S3: 5-10 (J)}} & \multicolumn{2}{c}{\textbf{S4: MIX (C)}}\\
        \cmidrule(lr){2-4} \cmidrule(lr){5-7} \cmidrule(lr){8-10} \cmidrule(lr){11-12}
        & R-FF & R-nFF & U & R-FF & R-nFF & U & R-FF & R-nFF & U & R & U\\
        \midrule
        Case 6  & -  & -  & -  & -  & 13 & 21 & -  & 37 & 55 & -  & -  \\
        
        Case 7  & -  & 35 & 28 & -  & 22 & 16 & -  & 25 & 14 & -  & -  \\
        
        Case 8  & 33 & 49 & 57 & 27 & 43 & 51 & 48 & 45 & 59 & -  & -  \\
        
        Case 9  & 11 & -  & -  & 17 & -  & -  & 19 & -  & -  & -  & -  \\

        Case 10 & 11 & 10 & 18 & 8  & 19 & 33 & 7  & 15 & 26 & -  & -  \\

        Case 11 & -  & -  & -  & -  & 0  & 0  & -  & 5  & 12  & -  & -  \\

        Case 12 & -  & -  & -  & -  & 3  & 3  & -  & 2  & 7  & -  & -  \\

        Case 14 & -  & -  & -  & -  & -  & -  & -  & -  & -  & 26 & 21 \\

        Case 15 & -  & -  & -  & -  & -  & -  & -  & -  & -  & 47 & 51 \\

        Case 16 & -  & -  & -  & -  & -  & -  & -  & -  & -  & 31 & 22 \\
        
        \bottomrule[1.5pt]
    \end{tabular}
\label{tab:case_count_realistic}
}
\end{table}

\Cref{fig:C_result} presents the live behavior inference results for C++-based routers in \textbf{S4}. The coarse-grained method, limited to approximating session boundaries, yields substantial inference biases. Incorporating the join and leave behavior identification methods significantly improves accuracy, reducing the median join time bias to under 5 seconds and the median leave time bias to below 10 seconds. Although observable leave behavior occasionally fails to appear---resulting in a few outliers---the leave behavior complement method effectively bounds all leave time biases within 15 minutes.

The above results indicate that when we have complete data, our live behavior inference method maintains a high level of consistency with the ground truth across various scenarios.

\subsubsection{Reliability (\textit{RQ2})}
\label{subsubsec:exp_reliability}

\begin{table}[!ht]
\centering
\scriptsize{
    \caption{Upper quartile value of inference biases in join and leave times for various router categories, with and without session complement method, and the corresponding similarity distance to the ground truth per hidden service's online session.}

    \setlength{\tabcolsep}{1.8pt}
    \begin{tabular}{M{0.75cm} rrr rrr rrr rrr rrr}
        \toprule[1.5pt]
         \multirow{2}{=}{\textbf{Compl\-ement}} & \multicolumn{3}{c}{\textbf{J-R-FF}} & \multicolumn{3}{c}{\textbf{J-R-nFF}} & \multicolumn{3}{c}{\textbf{J-U}} & \multicolumn{3}{c}{\textbf{C-R}} & \multicolumn{3}{c}{\textbf{C-U}} \\
        \cmidrule(lr){2-4} \cmidrule(lr){5-7} \cmidrule(lr){8-10} \cmidrule(lr){11-13} \cmidrule(lr){14-16}
         & join & leave & $\tau$ & join & leave & $\tau$ & join & leave & $\tau$ & join & leave & $\tau$ & join & leave & $\tau$ \\
        \midrule
        
        \ding{55} & 353 & 1147 & 25 & 1677 & 2336 & 66 & 2652 & 3722 & 106 & 868 & 2380 & 54 & 867 & 542 & 23 \\
        
        \ding{51} & 6 & 606 & 10 & 361 & 702 & 18 & 561 & 1436 & 33 & 2 & 430& 7 & 3 & 162 & 3 \\ 
        
        \bottomrule[1.5pt]
    \end{tabular}
\label{tab:exp_result}
}
\vspace{-3mm}
\end{table}

\Cref{tab:case_count_realistic} presents the occurrences of incorrect inferences in the fine-grained online session inference method caused by missing \textit{RouterInfo}, as listed in \Cref{tab:special_cases}, using the evaluation dataset collected from controlled floodfill routers. All cases occur, though with varying frequencies. Cases 6–10, and 14–16 appear more often and are successfully corrected by our online session complement methods. We further analyze the two unresolvable cases: 11 and 12.
\textbf{Case 11} is observed to be extremely rare. Although it could theoretically arise in \textbf{S2} for non-floodfill routers with reachable ports, no such instance is observed in practice.  Similarly, while \textbf{Case 12} may substantially affect inference accuracy, its occurrence is also, as expected, exceedingly rare. Then we quantify the effectiveness of our online session complement method by comparing the accuracy of inferred join and leave times with and without the session complement method.  \looseness=-1

To evaluate the effectiveness of the online session complement method, we quantitatively compare the live behavior inference biases across different router types and scenarios, using the fine-grained inference method alone and with the complement method, based on the evaluation dataset. The results are presented in \Cref{fig:incomplete_data_results}.

\Cref{fig:45-60_recovery} shows the inference biases of controlled host routers’ join and leave times in \textbf{S1}. Java-based floodfill routers, which exhibit explicit join and leave behaviors, are minimally affected by missing \textit{RouterInfo}, with median inference biases for both join and leave times below 10 seconds. In contrast, Java-based non-floodfill routers are more vulnerable to data loss. When using only the fine-grained method, the median leave time bias reaches 1400 seconds, while the online session complement method reduces it to 600 seconds. For Java-based unreachable routers, limited connectivity hampers \textit{RouterInfo} propagation compared to reachable routers, resulting in more missing \textit{RouterInfos}. Consequently, the fine-grained method produces median biases of 1350 seconds for join times and 2600 seconds for leave times, which the complement method effectively reduces to 600 and 1700 seconds, respectively.

\Cref{fig:30-45_recovery} presents the performance of the online session complement method in \textbf{S2}. The method significantly improves inference accuracy for Java-based floodfill routers, achieving median biases of less than 10 seconds for both join and leave times.
For non-floodfill routers, since multiple cases occur more frequently than in floodfill routers (as shown in \Cref{tab:case_count_realistic}), larger inference errors are observed. Regarding routers’ join times, the online session complement method decreases the median bias from 1130 seconds to 230 seconds. For leave times, although the complement method slightly raises the median bias, it effectively eliminates most outliers by correcting the majority of online session misidentifications, thereby concentrating the inference biases within a lower range of 600–1500 seconds. Finally, for unreachable routers, the online session complement method is also highly effective, reducing the median inference biases for join times from 1250 seconds to within 10 seconds, and for leave times from 1300 to 800 seconds.

\Cref{fig:5-10_recovery} shows that, in \textbf{S3}, the online session complement approach significantly enhances the accuracy of join and leave time inference for Java-based routers. Along with the results in \Cref{fig:45-60_recovery} and \Cref{fig:30-45_recovery}, these findings demonstrate the robustness of \sys in accurately inferring the live behaviors of Java-based routers under real-world conditions.

\Cref{fig:C_recovery} illustrates the performance of our session-complement approach for C++-based routers in \textbf{S4}. For reachable routers, the method reduces all join time inference biases to within 10 seconds and eliminates most inference biases when inferring leave times, keeping 75\% of values below 550 seconds. For unreachable routers, join time inference accuracy is comparable to that of reachable ones. The leave time inference shows even greater improvement, with nearly all inference biases constrained within 240 seconds.

\begin{figure}
    \centering
    \includegraphics[width=0.85\linewidth]{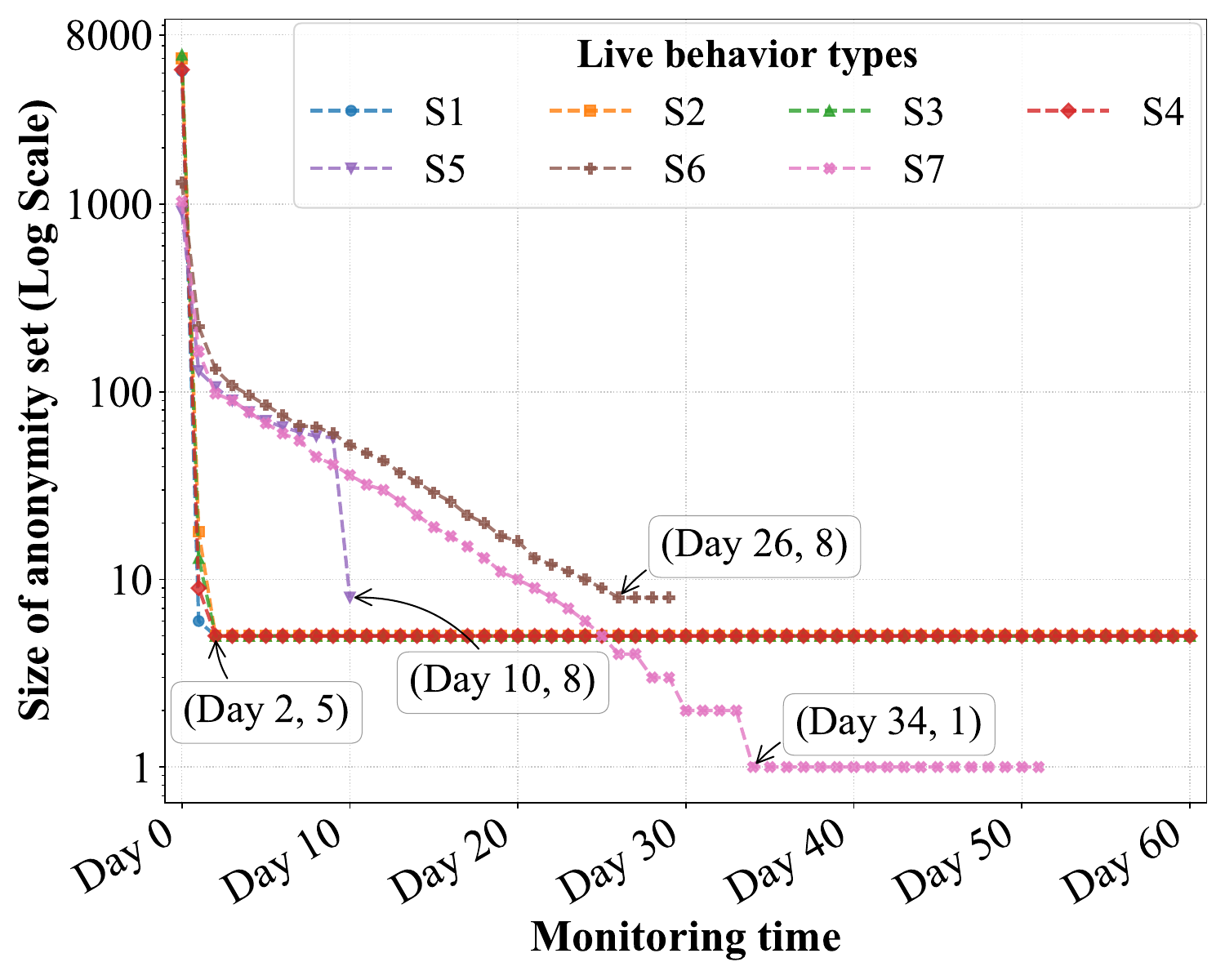}
    \caption{Hidden services' anonymity set size vs. time. Note: When a scenario contains only a single service and no other synchronized services, that service can be uniquely identified: S1, S2, S3, and S4 on Day 2; S5 on Day 10; S6 on Day 26; and S7 on Day 34.}
    \label{fig:anonimityVStime}
    \vspace{-5mm}
\end{figure}

\Cref{tab:exp_result} summarizes the inference accuracy of join and leave times for different types of I2P routers across all scenarios, comparing the results of using the fine-grained online session inference method alone and in combination with the online session complement method. In the table, each `join' column represents the upper quartile of biases (i.e., 75\% of biases are less than or equal to this value) for all inferred join times of the corresponding router type. Likewise, the “leave” column indicates the upper quartile of leave time inference biases. These results highlight that the session complement method substantially enhances the inference accuracy of router live behaviors in real-world conditions.

Based on the join and leave time inference biases discussed above, we derive the similarity thresholds used to partition anonymity sets in deanonymization attacks against hidden services in the real-world environment. The experimental results indicate that inference biases vary across different router types. Accordingly, for each online session (i.e., one on–off cycle) of a target hidden service, we assign distinct similarity thresholds to different router categories to determine whether a router belongs to the service’s anonymity set. \textit{Thr}\textsubscript{DTW} in \Cref{tab:exp_result} denotes the per-session similarity threshold for each router type, derived from its corresponding live behavior inference accuracy. 
Consequently, the overall threshold for determining a router’s inclusion in the anonymity set is defined as $n \times thr_c$, where $n$ is the number of the hidden service’s online sessions and $thr_c$ is the category-specific threshold.

\subsubsection{Effectiveness (\textit{RQ3})}
\label{subsubsec:Effectiveness}

To evaluate the real-world performance of \sys, we conduct deanonymization attacks on controlled hidden services under the scenarios defined in \Cref{tab:behavior_setting}. For each scenario \textbf{S1}--\textbf{S4}, five hidden services are configured to follow the same live behavior described in \S\ref{subsec:exp_setup}. For the long-lived scenarios \textbf{S5} and \textbf{S6}, eight services are deployed for each configuration, while one service is run for \textbf{S7}. Throughout their runtime, we continuously monitor the I2P network and record the size of each hidden service’s anonymity set at the end of each day. Please note: In practice, it is unlikely for two hidden services to exhibit the same on-off pattern, and a real-world hidden service will be uniquely identified. We create synchronized processes to illustrate \Cref{corollary::SynchronizedProcesses}. \looseness=-1

\Cref{fig:anonimityVStime} presents how the anonymity sets of controlled hidden services shrink over time. For the short-lived hidden services (\textbf{S1}--\textbf{S4}), their anonymity sets rapidly shrink from 5,000-6,000 I2P routers to approximately 10 on the first day. By day 2 and onward, their anonymity sets shrink to 5, corresponding to the number of synchronized services.
In the long-lived scenarios \textbf{S5}–\textbf{S7}, 915, 1,307, and 1,035 I2P routers go online around the same time. These routers form their initial anonymity sets. After one day, these sets shrink to 129, 223, and 164 routers. In \textbf{S5}, the set further decreases to 57 routers by day 9 and to 8 by day 10, corresponding to the number of synchronized services. In \textbf{S6}, routers in the initial set gradually go offline, reducing the anonymity set to 8 by day 26. Finally, the single hidden service in \textbf{S7} is successfully deanonymized after 34 days of monitoring.\looseness=-1

\begin{figure}[!ht]
    \vspace{-3mm}
    \centering

    \subfloat[]{\includegraphics[width=0.485\linewidth]{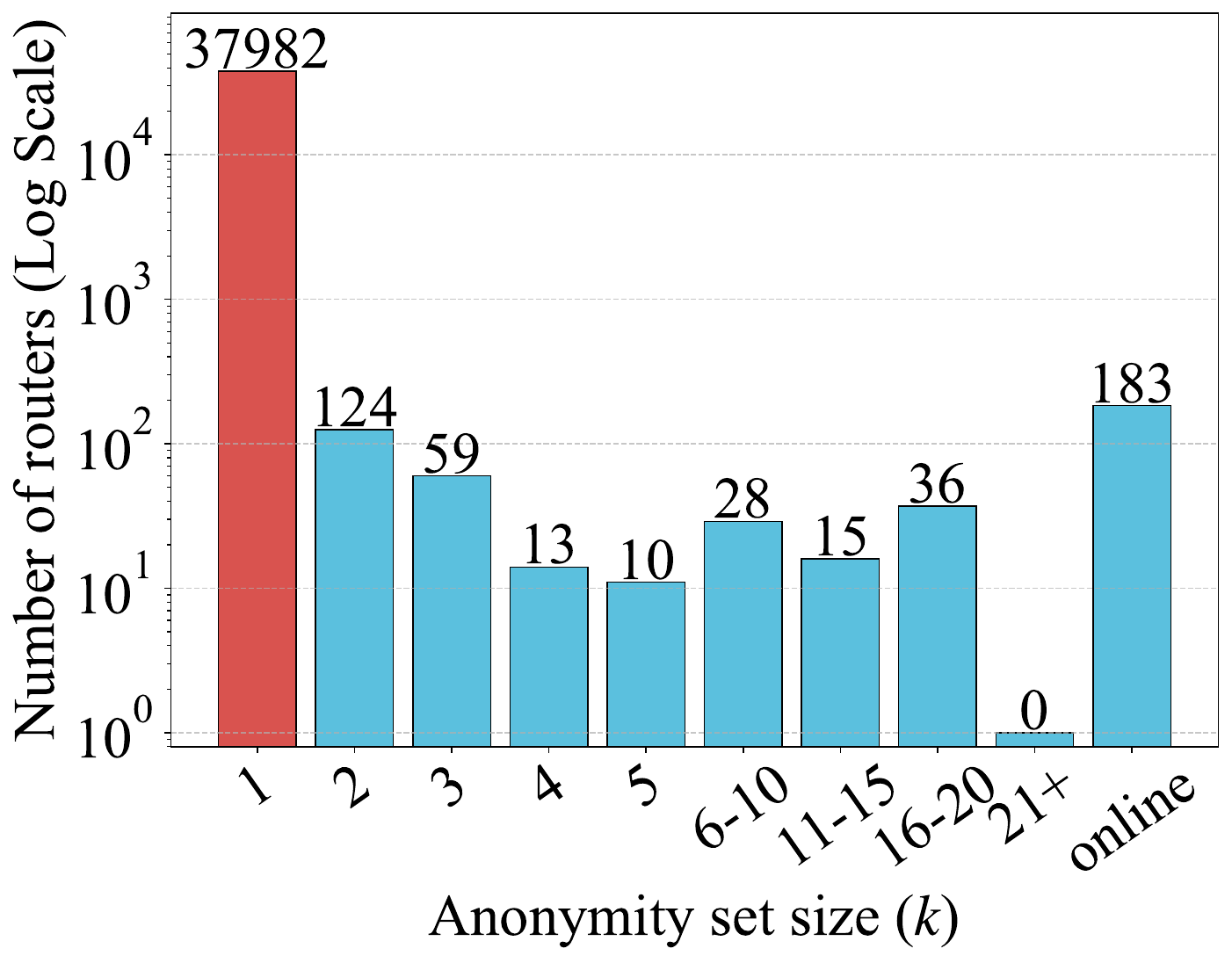}
    \label{fig:8-month-distinguishability}}
    \hfill
    \subfloat[]{\includegraphics[width=0.485\linewidth]{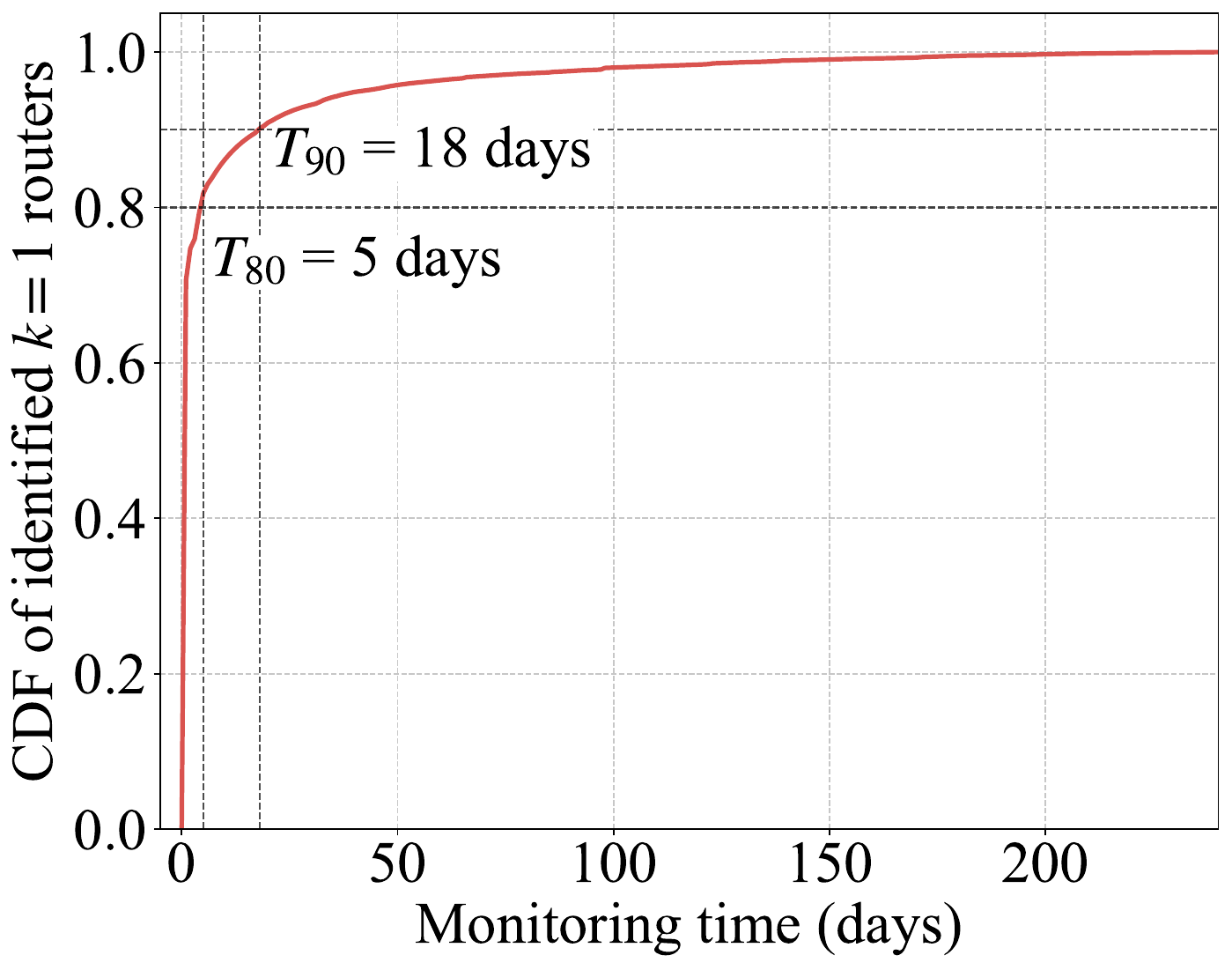}
    \label{fig:distinguish_efficiency}}
    
    \caption{Experimental results from long-term monitoring of the I2P network. (a) Number of routers with different anonymity set sizes (k). (b) CDF of the number of routers exhibiting unique behaviors over the monitoring period.}
    \label{fig:router_distinguishability}
    \vspace{-4mm}
\end{figure}

We also evaluate the behavioral distinguishability of real-world I2P routers by examining how many exhibit on–off patterns that are unique with respect to our live behavior inference framework and a predefined similarity threshold (Refer to \Cref{appendix:thr-discuss} for the choice of the threshold). We include all routers that appeared online on the first monitoring day and track their behaviors for the subsequent eight months. Each day, we compute the similarity distance between every router and all others to derive its anonymity set. A router is classified as behaviorally unique once no other router remains in its anonymity set. \Cref{fig:8-month-distinguishability} summarizes the results. Among all monitored routers, 37,982 eventually exhibit unique behaviors, while 468 share a similar behavior with at least one other router. \Cref{fig:distinguish_efficiency} further shows how quickly uniqueness emerges: 90\% of the unique routers become distinguishable within the first 18 days. These findings highlight the substantial behavioral diversity among I2P users, which implies that hidden services hosted on such routers may produce similarly distinguishable behavioral patterns, potentially increasing their deanonymization risk. For routers without unique behaviors, our method may not deanonymize them immediately, but it significantly reduces their anonymity set from thousands to tens or hundreds, enabling strong adversaries to focus active attacks (e.g., DoS, watermarking, fingerprinting) on a much smaller candidate pool. \looseness=-1

\section{Mitigation}
\label{sec:mitigation}
\looseness=-1

To ensure a comprehensive treatment of the security implications of our findings, we also analyze potential countermeasures to our attack. Mitigation can be considered from two complementary perspectives: \textbf{(I) protocol-level defenses} that reduce information leakage from RouterInfo publication, and \textbf{(II) user-side measures} that make hidden services harder to correlate. \looseness=-1

\textbf{Protocol-Level Defenses.} We propose two protocol-level strategies. \textbf{(a) Randomizing Publication Intervals:} We can enhance the randomness of \textit{RouterInfo} publication by substituting the fixed four update interval publication rule (as introduced in \S\ref{subsec:time_state}) with a randomized mechanism that sets the next mandatory publication within 10–55 minutes after each one. This broad randomization window prevents attackers from identifying online sessions through periodic patterns and, in real-world environments, removes the analytical basis for handling uncaptured \textit{RouterInfo}. Importantly, this approach still guarantees timely \textit{RouterInfo} refreshing in the netDB, since the maximum publication interval remains shorter than the \textit{RouterInfo} timeout threshold (i.e., one hour),making this mitigation strategy both secure and operationally adaptable. \textbf{(b) Eliminate Join and Leave behaviors:} Eliminating special \textit{RouterInfo} publication patterns during router initialization and shutdown phases can prevent attackers from precisely identifying the start and end of online sessions. Also, without these behaviors, a real-world attacker cannot determine whether the first \textit{RouterInfo} within a \textit{RouterInfo} trace is the initial one published after a router starts or if previous \textit{RouterInfos} were lost. This makes it infeasible to recover complete online sessions when only partial \textit{RouterInfo} data is available. \looseness=-1

\begin{importantblock}
\noindent\textbf{Real-World Deployment by the I2P Team.}
Following our responsible disclosure, I2P provides a patch to mitigate the anonymity risk from three perspectives. First, to introduce randomization, the patch dictates that each routine \textit{RouterInfo} publication has a 1-in-32 chance of being skipped. Second, the patch refines the management of the update counter and removes the specific code branch executed for the initial \textit{RouterInfo} publication, effectively eliminating the `join behavior' outlined in \S\ref{subsec:time_state}. Finally, a floodfill router now does not publish a non-floodfill \textit{RouterInfo} during a non-graceful shutdown, thus removing the `leave behavior'. \looseness=-1
\end{importantblock}

We also propose two user-side measures: \textbf{(a) Synchronizing the behavior of a group of I2P hidden services.} By coordinating several hidden services to exhibit identical behaviors, the host routers of these services can collectively form a larger anonymity set, preventing attackers from associating a single hidden service with a unique router. Although this strategy can effectively expand the anonymity set, it is generally impractical because hidden services operated by different users typically act independently. \textbf{(b) Detecting active probing of I2P services.} Although the hidden service cannot observe a client's IP address, it can observe the client's I2P identity. Repeated or periodic probing from the same identity, or probing with distinguishable timing patterns from different sources may reveal attempts to infer live behavior, allowing the service to ignore or throttle such requests. \looseness=-1

\section{Related Work}
\label{sec:related_work}

Over the years, numerous anonymous communication systems have been developed, such as Tor \cite{dingledine2004tor}, I2P \cite{zantout2011i2p}, to provide users with anonymous access to Internet services. Meanwhile, various attacks have been investigated to break the users' anonymity. Broadly speaking, the core idea of deanonymization is to exploit identifiable behavioral patterns over time to gradually reduce the anonymity set of a target user, ultimately isolating the target from a crowd of users. \looseness=-1

In the I2P network, researchers have applied deanonymization strategies by leveraging the network’s unique protocol behaviors and inherent vulnerabilities. Herrmann \emph{et al.} \cite{herrmann2011privacy} attempted to trace hidden services by correlating traffic observed on controlled routers participating in the service's tunnels. Other attacks have targeted specific protocol vulnerabilities. Egger \emph{et al.} \cite{egger2013practical}, for instance, launched a Sybil attack against hidden services. Their method involved using numerous floodfill routers to intercept the published \textit{RouterInfo} of a service's host router, thereby revealing the identity of the associated tunnel's exit router. In another approach, Crenshaw \emph{et al.} \cite{crenshaw2011darknets} proposed a side-channel method that correlates anonymous I2P websites (eepsites) with their public-facing counterparts on the clearnet to uncover the server's true IP address. Although prior studies have demonstrated successful deanonymization, they either exploited protocol vulnerabilities in I2P or relied on traffic-correlation techniques. The protocol vulnerabilities leveraged by previous work have been patched and are no longer exploitable. Moreover, traffic-correlation attacks require the adversary to occupy specific positions within a hidden service’s client tunnel, which is extremely difficult to achieve in the I2P network at its current scale.\looseness=-1  

More recently, research has shifted toward analyzing behavioral patterns of I2P routers. Simioni \emph{et al}. \cite{simioni2021monitoring} inferred a router’s online and offline status based on the duration its \textit{RouterInfo} remained in the netDB, thereby identifying behavioral differences among routers. However, their work does not uncover the \textit{RouterInfo} publication mechanism and, as a result, cannot achieve high accuracy when inferring I2P routers' live behaviors.
Moreover, in real-world scenarios where capturing all \textit{RouterInfos} is impossible, their method cannot handle the impact of \textit{RouterInfo} loss, thus making it unable to achieve reliable deanonymization.\looseness=-1

Previous researchers on Tor have adopted different types of identifiable patterns, ranging from embedding actively crafted traffic features into flows of anonymous systems \cite{ling_new_2009,murdoch_low-cost_2005,wang_network_2007,mittal_stealthy_2011,biryukov_trawling_2013,houmansadr_swirl_2011,borisov_denial_2007,chakravarty_traffic_2010,evans_practical_2009}, mining intrinsic traffic characteristics of anonymity systems \cite{wacek_empirical_2013,edman_as-awareness_2009,matic_caronte_2015,kwon_circuit_2015,rochet_claps_2020,sun_counter-raptor_2017,oh_deepcoffea_2022,nasr_deepcorr_2018,berthold_dummy_2003,lopes_flow_2024,overlier_locating_2006,bauer_low-resource_2007,nithyanand_measuring_2015,backes_nothing_2014,sun_raptor_2015,shmatikov_timing_2006,johnson_users_2013,rochet_waterfilling_2017, Zhang_Paxson_2000,Blum_Song_Venkataraman_2004,mathewson2004practical,danezis_two-sided_2007,murdoch_sampled_2007}, detecting side-channel behaviors such as clock skew on the host of a target hidden service \cite{murdoch2006hot}, to identifying protocol flaws that cause behavioral deviations across anonymous users \cite{jansen2014sniper}.
Existing deanonymization methods for Tor hidden services do not leverage user-driven behavioral patterns that produce distinctive on–off activity in hidden services.\looseness=-1

\section{Conclusion}
\label{sec:conclusion}

We propose \sys, a de-anonymization framework for I2P hidden service. Using the routers' live behaviors inferred by a passive monitoring approach to correlate with the service's live behavior obtained via an active probing approach, we may identify the server's host router to reveal the IP address of the hidden server. To precisely infer the I2P router's fine-grained live behavior, we explore the publication rules of \textit{RouterInfo} data during join and leave period to precisely identify the router's join and leave behavior. Also, we investigate the online session recovery method to reduce the impact of the incomplete data caused by low-cost data collection strategy. The real-world experiment results shows that the live behavior inferred by \sys has high consistency with the router's actual live behavior. Moreover, all the hidden services we deployed were successfully deanonymized by \sys.

\section*{Acknowledgment}
\label{subsec:acknowledge}

We thank the anonymous reviewers for their constructive comments and suggestions. This work is supported in part by the National Natural Science Foundation of China (NSFC) under Grant Nos. 62232004, 92467205, and 62502086, the Natural Science Foundation of Jiangsu Province under Grant No. BK20251295, the Start-up Research Fund of Southeast University under Grant No. RF1028624178, the Jiangsu Provincial Key Laboratory of Network and Information Security under Grant No. BM2003201, the Key Laboratory of Computer Network and Information Integration of Ministry of Education of China under Grant No. 93K-9, and the Collaborative Innovation Center of Novel Software Technology and Industrialization. We also acknowledge the support of the Big Data Computing Center of Southeast University. Any opinions, findings, conclusions, and recommendations in this paper are those of the authors and do not necessarily reflect the views of the funding agencies.

\section*{Ethics Considerations}
\label{subsec:ethic}

In this paper, we collected I2P router's \textit{RouterInfo} data, and periodically sniffed several controlled I2P hidden services, aiming to correlate the live behavior of the target hidden services with their host routers, so as to achieve de-anonymization. We took the following steps to ensure that our experiments were conducted ethically:

\textbf{Responsible Disclosure}: Upon discovering the vulnerabilities, we disclosed our findings to the I2P maintainers to support timely mitigation. Our work was acknowledged by the I2P project, and as of this submission, the reported issues have been addressed in a prior I2P release. \cite{i2pRepairedRelease}.

\textbf{Controlled Deployment}: All hidden services considered as attack targets are deployed on controlled routers by ourselves, and the identities of these services are not disclosed to anyone, ensuring that no user could accidentally access these hidden services.

\textbf{I2P Academic Research Guidance}: Our experiments strictly adhere to the security recommendations on the I2P Academic Research website~\cite{I2P_VRP}. Specifically, we ensure the following: no active exploits or Denial of Service (DoS) attacks are performed on the I2P network; no social engineering is conducted targeting I2P team members or community participants; and no physical or electronic attempts are made against I2P property or data centers.

\textbf{Data Protection}: We implemented extensive measures to minimize potential harm resulting from our data collection efforts, including the following: (i) All data used in our experiments was collected from 15 floodfill routers that we deployed. These routers participated in maintaining the netDB and constructing tunnels for other users, ensuring that the normal operation of the I2P anonymous network was not disrupted. Furthermore, all data analysis was conducted offline to avoid any impact on other users. (ii) We ensured the confidentiality of all user data and metadata, with no information disclosed to third parties. (iii) The collected data was securely transmitted via SSH-encrypted channels and stored on a secure server within a campus machine room with restricted physical access. (iv) Our interaction with the data was limited to observation, with no modifications made. All collected data was anonymized and subsequently deleted following the submission of this paper.

\bibliographystyle{IEEEtran}
\bibliography{reference_new}

\appendices
\crefalias{section}{appendix}
\crefalias{subsection}{appendix}
\renewcommand{\thesubsection}{\thesection-\Alph{subsection}}

\section{Low-cost deployment strategy for floodfill routers}
\label{subsec:floodfill_num}

\begin{figure}[!th]
    \centering
    \subfloat[]{\includegraphics[width=0.485\linewidth]{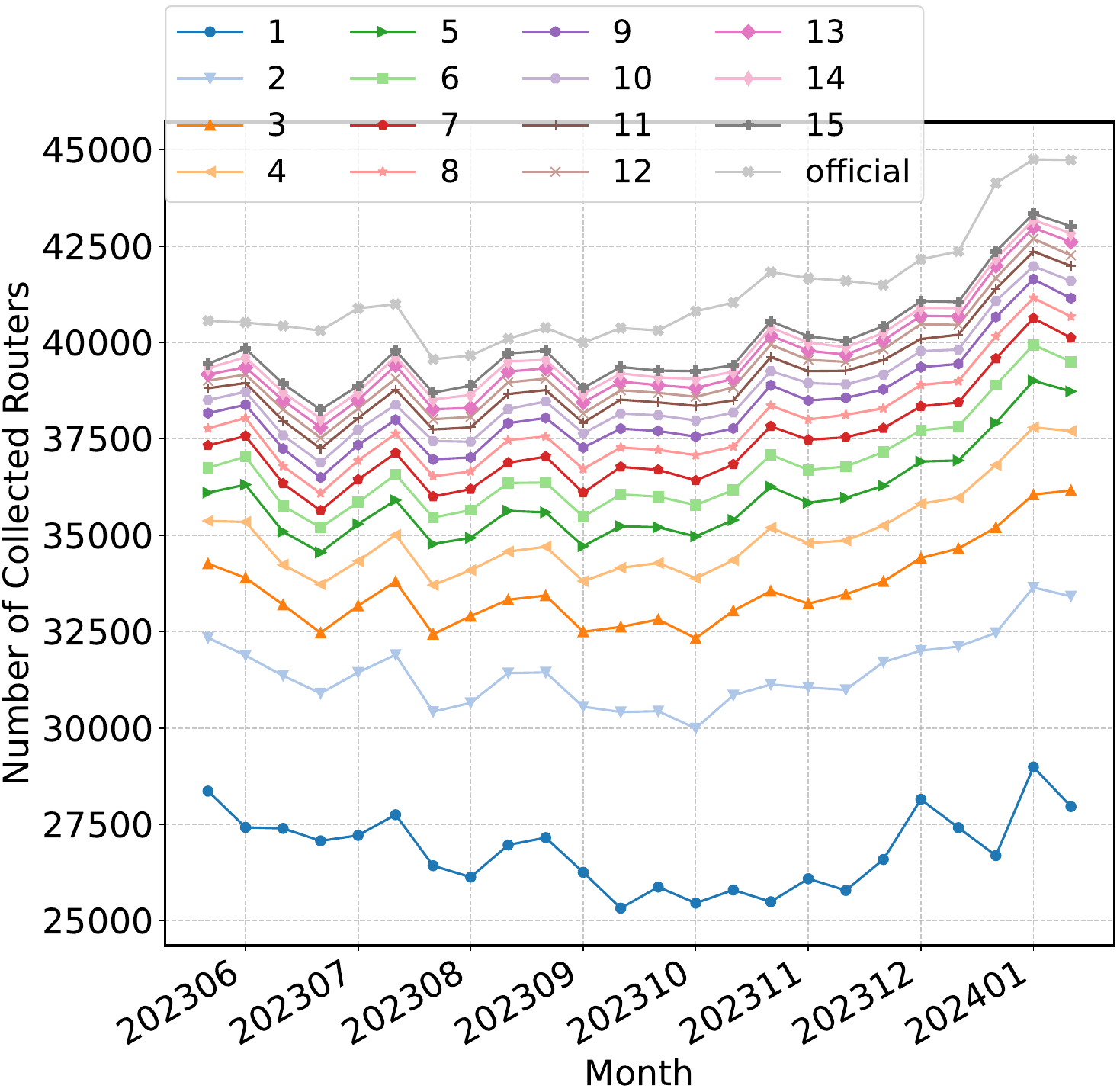}
    \label{fig:network_coverage}}
    \hfill
    \subfloat[]{\includegraphics[width=0.485\linewidth]{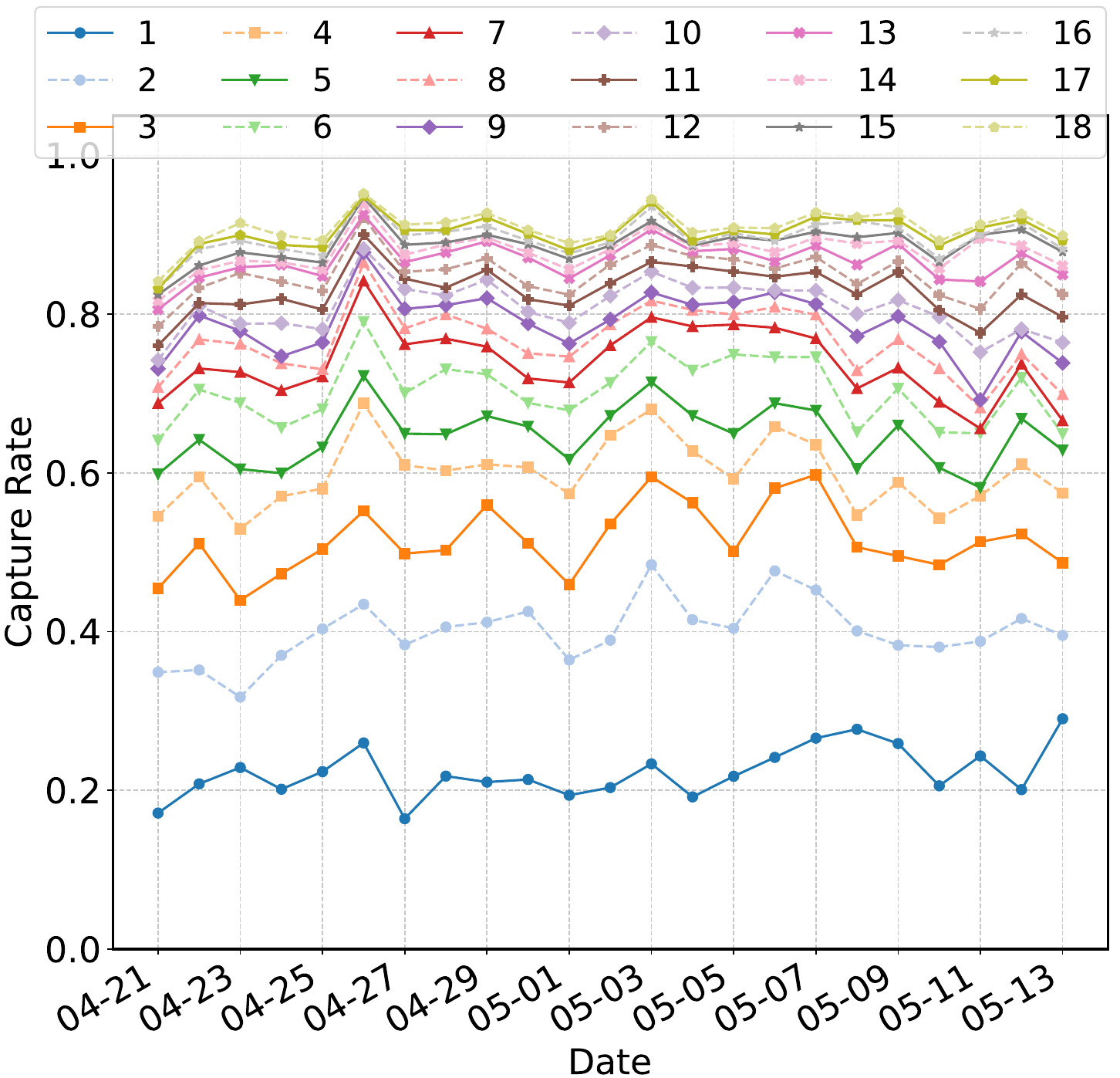}
    \label{fig:ri_coverage}}
    
    \caption{Floodfill routers' effectiveness. (a) Comparison between our collected data and official data. (b) RouterInfo Capture Rate vs. Number of Controlled Routers.}
    \label{fig:floodfillEffectiveness}
\end{figure}

We evaluate the data collection effectiveness of 15 floodfill routers using two key metrics: the network-wide router coverage and the \textit{RouterInfo} capture rate from controlled routers. Over an 8-month monitoring period (2023.06–2024.02), we continuously recorded the number of unique I2P routers detected each day and compared it with the official statistics \cite{i2pmetrics} to validate our network coverage, as shown in \Cref{fig:network_coverage}. With 15 floodfill routers, our system achieved 98\% coverage of the officially reported router count, with only marginal gains observed beyond 13 routers. We further assessed the capture rate of \textit{RouterInfo} data by comparing the records collected by our floodfill routers with the actual number of \textit{RouterInfo} publications logged locally by the controlled host routers. \Cref{fig:ri_coverage} presents the average capture rates achieved using different numbers of floodfill routers. A 15-router deployment yields a 90\% capture rate, comparable to that of 18 routers. Therefore, we use 15 floodfill routers in all real-world experiments to ensure reliable data collection with minimal resource overhead.

\section{Supplementary content for \sys}
\label{sec:additional_tech}

\subsection{Online session inference for C++-based routers}
\label{subsubsec:cpp_code_analysis}

\textbf{Coarse-grained online session inference for C++-based router.} For C++-based routers, we identify online sessions by leveraging the periodic
congestion-evaluation task, which executes every 12 minutes. Whenever the
measured congestion level differs from the previous value, the router publishes a new \textit{RouterInfo}. Such publications serve as routine \textit{RouterInfos}. Consequently, the interval between two routine \textit{RouterInfos} must be a multiple of 12 minutes. By scanning the \textit{RouterInfo} trace for changes in the congestion flag within the \textit{options} field, we identify the first routine \textit{RouterInfo} and subsequently detect all following routine \textit{RouterInfos}. If two successive routine \textit{RouterInfos} fail to satisfy the 12-minute interval constraint, we infer that the router temporarily went offline, thus marking a session boundary. Since congestion levels vary frequently due to dynamic tunnel participation, routine \textit{RouterInfos} are published regularly, providing sufficient granularity for reliable session inference. \looseness=-1

Unlike the coarse-grained method for Java-based routers, the method for C++-based routers relies on tasks with strict timing constraints (i.e., 12-minute intervals), resulting in high reliability for online session identification. If the router briefly goes offline and then returns online, the coarse-grained method can robustly identify two separate sessions based on the irregular intervals between the last routine \textit{RouterInfo} in the preceding session and the first routine \textit{RouterInfo} in the succeeding session. However, between the two routine \textit{RouterInfo} data, there might exist several non-routine \textit{RouterInfo} published by other tasks. It remains unclear which session these non-routine \textit{RouterInfo} data belong to, making it impossible to accurately determine the router’s join and leave times. To address this limitation, we introduce a join behavior identification method. \looseness=-1

\textbf{Join behavior identification for C++-based routers.} After the router's startup, the router sets a timer for the next congestion level evaluation, and then the initial \textit{RouterInfo} is published within around 500 ms. Consequently, the interval between the initial \textit{RouterInfo} data and the first routine \textit{RouterInfo} is also expected to be a multiple of 12 minutes minus 500 ms. Based on this join behavior, we could search the \textit{RouterInfo} trace before the first routine \textit{RouterInfo} to identify the initial \textit{RouterInfo}. The publication time of this initial \textit{RouterInfo} is designated as the start time of the online session. After detecting the join behavior, all \textit{RouterInfos} before the online session's initial \textit{RouterInfo} belong to the preceding online session. \looseness=-1

However, if the router reconnects and remains online only briefly after a temporary offline period, resulting in no routine \textit{RouterInfo} being published during the subsequent session (see \textbf{Case 13} in \Cref{tab:special_cases}), rendering the join behavior incomplete and the identification method inapplicable. To address this, we provide the leave behavior identification method. \looseness=-1

\textbf{Leave behavior identification for C++-based routers.} 
C++-based routers can shut down either gracefully or abruptly. In a graceful shutdown, the router notifies peers and completes ongoing transfers, taking up to 10 minutes. If a congestion level evaluation occurs during the shutdown, the router publishes a \textit{RouterInfo} with a `G' congestion flag, indicating severe congestion and an inability to accept new connections. This \textit{RouterInfo} with the `G' flag is identified as a final publication, marking the end of the online session. Any \textit{RouterInfo} after this is part of the succeeding session. Otherwise, if no congestion level evaluation occurs within the 10-minute window, no \textit{RouterInfo} with the G' flag will be published, and the leave behavior will be missing. Additionally, in the case of an abrupt shutdown, the router halts all operations immediately, potentially interrupting data transfers and dropping connections without notification. This case also lacks any explicit leave indicator. \looseness=-1

\subsection{I2P Software Implementation Identification}
\label{subsubsec:versions}

We distinguish the implementation of I2P software (i.e., Java or C++) by analyzing their \textit{RouterInfo} data. The primary differences in their \textit{RouterInfo} data are found within the \textit{RouterAddress} field, as illustrated in \Cref{tab:language_diff}. Starting with version 0.9.55, the C++ implementation deprecates the SSU protocol and exclusively supports the SSU2 protocol (an updated version of SSU), while the Java implementation removed SSU support starting in version 0.9.61. Additionally, the \textit{cost} field values differ: Java assigns a value between 10 and 12 for the \textit{RouterAddress} using NTCP2 and 4 to 8 for those using SSU or SSU2, while C++ assigns 3 for the \textit{RouterAddress} using NTCP2 and 8 for those using SSU2. By examining these fields, we can distinguish between the two implementations. \looseness=-1

\begin{table}[th]
    \centering
    \caption{Different implementation between Java and C++.}
    \label{tab:language_diff}
    \begin{threeparttable}
    \scriptsize
    \begin{tabularx}{0.95\linewidth}{c X X X}
        \toprule[1.5pt]
        \textbf{Implementation} & \textbf{FS} & \textbf{CN} & \textbf{CS} \\
        \midrule
        Java & 0.9.61 & 10 -- 12 & 4 -- 8 \\
        C++ & 0.9.55 & 3 & 8 \\
        \bottomrule[1.5pt]
    \end{tabularx}
    \begin{tablenotes}
        \footnotesize 
        \item \textbf{FS}: First SSU2-Supported version. \textbf{CN}: NTCP2 Address Cost. \textbf{CS}: SSU/SSU2 Address Cost.
    \end{tablenotes}
    \end{threeparttable}
    \vspace{-3mm}
\end{table}

\subsection{Online session complement for C++-based routers}
\label{subsec:C++-based_complement}

\textbf{Solution 1: Online session concatenation.} For C++ routers, the loss of a routine \textit{RouterInfo} can cause the next non-routine \textit{RouterInfo} to be mistakenly identified as routine if it has a different congestion flag (\textbf{Case 14} in \Cref{tab:special_cases}). In such cases, the routine \textit{RouterInfo} cannot satisfy the 12-minute interval requirement, leading to the split of a single session into three. We address this by verifying if the time interval between the first and third routine \textit{RouterInfo} meets the 12-minute multiple interval condition, allowing the sessions to be merged if the condition is satisfied. \looseness=-1

\textbf{Solution 2: Join behavior supplementation.} To supplement the first \textit{RouterInfo} data in an online session of a C++-based router (\textbf{Case 15} in \Cref{tab:special_cases}), we utilize a specific type of \textit{RouterInfo} known as the peer testing \textit{RouterInfo}, generated by a periodic peer testing task that evaluates the router’s port reachability. This task runs every 71 minutes and publishes the constructed \textit{RouterInfo} to the netDB. Therefore, if the time interval between two \textit{RouterInfo} data is 71 minutes, they are identified as peer testing \textit{RouterInfo}. Similar to the congestion level evaluation task for generating routine \textit{RouterInfo}, the timer for the peer testing task is set immediately after the router's startup. Consequently, the interval between the first \textit{RouterInfo} data and a peer testing \textit{RouterInfo} should be a multiple of 71 minutes, while the interval with routine \textit{RouterInfo} should be a multiple of 12 minutes. \looseness=-1

Using this insight, we identify the first peer testing \textit{RouterInfo} within a session, then search backward for a time point satisfying both constraints: (1) the interval to the peer testing \textit{RouterInfo} is a multiple of 71 minutes, and (2) the interval to the first routine \textit{RouterInfo} is a multiple of 12 minutes. This time point is when the first \textit{RouterInfo} is expected to be published. This determination method is theoretically highly accurate, as the next matching that satisfies the interval requirement occur before $71 \times 12 = 852$ minutes, over 14 hours. An error in judgment would only arise if the router remained online for more than 14 hours without any \textit{RouterInfo} being captured. However, in subsequent experiments \S\ref{subsec:exp_setup}, we verify that this situation rarely occurs. \looseness=-1

\textbf{Solution 3: Leave behavior supplementation.} For C++-based routers, if no data has been published within 30 minutes, a router is forced to publish a new \textit{RouterInfo}. Therefore, the router is expected to go offline within 30 minutes after the last \textit{RouterInfo} publication. Also, the probability of the router going offline at any moment within this 30-minute window is equal. Therefore, we use 15 minutes as the expected remaining online duration from the publication time of the last \textit{RouterInfo} data. \looseness=-1

\subsection{Firewalled Java Router Handler}
\label{sec:firewalled_java}
When analyzing routers behind firewalls or NAT, our live behavior inference method for Java-based routers often encounter interference. We find that such routers frequently publish \textit{RouterInfo} entries with empty IP addresses in the \textit{RouterAddresses} field, indicating non-reachability. This section explains the underlying publication mechanism of firewalled routers and presents our approach to mitigate its impact. \looseness=-1

Routers behind firewalls assess port reachability via peer testing tasks with assistance from other peers. If the test reveals that the router can no longer receive messages from unknown peers, its IP addresses in all associated \textit{RouterAddresses} are set to null, and the port is set to 0. To notify the network of this status change, a new \textit{RouterInfo} is immediately generated and published to the netDB. \looseness=-1

Once identified as firewalled, the router must rely on introducers to accept incoming connections. It selects up to three introducers and establishes sessions with them. Upon success, another \textit{RouterInfo} is published, embedding the introducer information within the \textit{RouterAddresses} using the SSU protocol. However, as each introducer has an expiration time, the router must periodically refresh or replace them, depending on session availability. This dynamic process results in frequent, irregular \textit{RouterInfo} publications. These additional publications increase the likelihood that the periodic update task will be constrained by the 9-minute delay after the last successful publication, as specified in \S\ref{subsec:time_state}. If introducer updates trigger frequent non-routine \textit{RouterInfo} publications after a routine publication, the next routine \textit{RouterInfo} may be delayed beyond the expected interval. Consequently, our method for identifying routine \textit{RouterInfo} based on fixed timing may become unreliable. \looseness=-1

To overcome this challenge, we propose a method for identifying routine \textit{RouterInfo} in Java-based firewalled routers. We observe that routine update tasks do not alter introducer information. Thus, if introducer details remain unchanged between two \textit{RouterInfo}, the latter is likely a routine publication; otherwise, it is attributed to a peer testing task. Once routine \textit{RouterInfo} is identified, we can appropriately relax the interval constraints by considering the number of non-routine publications between them. Moreover, firewalled routers tend to exhibit a characteristic join behavior. Due to the delay in completing initial peer testing, the first \textit{RouterInfo} published often lacks both the `R' and `U' reachability flags, indicating the router’s reachability status remains undetermined at that moment.
\looseness=-1

\section{\texorpdfstring{Probability that no two processes are the same after $m$ cycles}{Probability that no two processes are the same after m cycles}}
\label{sec:ProbabilityAnalysis}

\subsection{Problem Definition}
Given $n$ on-off stochastic processes. Their $on$ periods follow the same discrete probability distribution and have $k$ values with probabilities $a_1, \dots, a_k$. Their $off$ periods follow the same discrete probability distribution and have $l$ values with probabilities $b_1, \dots, b_l$. The $n$ processes are independent and their on-off cycles are independent. One cycle contains one $on$ period and one $off$ period.

What is the probability $P_{\mathcal{A}}$ that one particular process $A$ does not produce the same length-m sequence with any of the other n-1 processes?

\subsection{Solution}

\textbf{Step 1. Model a single cycle.}
One cycle \( C \) is an ordered pair \( (O,F) \) where \( O \) is an on-period
chosen from \(\{1,\dots,k\}\) with probabilities \( a_1,\dots,a_k \),
and \( F \) is an off-period chosen from \(\{1,\dots,l\}\) with probabilities \( b_1,\dots,b_l \).
Since \( O \) and \( F \) are independent,
\begin{equation}
P(C=(i,j)) = a_i b_j,
\end{equation}
where $i = 1,\dots,k$ and $j = 1,\dots,l$. Note:
\begin{equation}
\sum_{i=1}^k a_i = 1, \qquad
\sum_{j=1}^l b_j = 1
\end{equation}

\textbf{Step 2. Probability $P_{\mathcal{C}}$ two independent processes match on one cycle.}
Let process \( A \) produce cycle \( C \) and an independent process \( B \) produce cycle \( C' \).
The probability they coincide on a particular cycle value \((i,j)\) is
\begin{equation}
P(C=(i,j))\,P(C'=(i,j)) = (a_i b_j)^2.
\end{equation}
Summing over all possible outcomes gives
\begin{equation}
P_{\mathcal{C}} = \sum_{i=1}^k \sum_{j=1}^l (a_i b_j)^2
= \left( \sum_{i=1}^k a_i^2 \right)
  \left( \sum_{j=1}^l b_j^2 \right).
\end{equation}
Define
\begin{equation}
S_a = \sum_{i=1}^k a_i^2, \qquad
S_b = \sum_{j=1}^l b_j^2.
\end{equation}
Therefore,
\begin{equation}
P_{\mathcal{C}} =S_a S_b.
\end{equation}

\textbf{Step 3. Probability $P_m$ that two processes of length-m match.}
A length-\( m \) sequence consists of \( m \) independent cycles.
Two processes produce the same sequence exactly when they match in all \( m \) cycles.
Since cycles are independent,
\begin{equation}
P_m = (S_a S_b)^m.
\end{equation}

\textbf{Step 4. Probability $P_{\mathcal{A}}$ that none of the other n-1 processes match process A}.
Fix the realized length-\( m \) sequence of process \( A \).
For any other process \( B \),
the probability it matches \( A \) on that fixed sequence is \( P_m=(S_a S_b)^m \).
Thus the probability it does \emph{not} match is \( 1 - P_m \).
Since all \( n-1 \) other processes are independent,
\begin{equation}
P_{\mathcal{A}}
= \big( 1 - P_m \big)^{n-1}.
\end{equation}

\section{Discussion on the similarity threshold for distinguishing router behaviors}
\label{appendix:thr-discuss}

\begin{figure}
    \centering
    \includegraphics[width=0.9\linewidth]{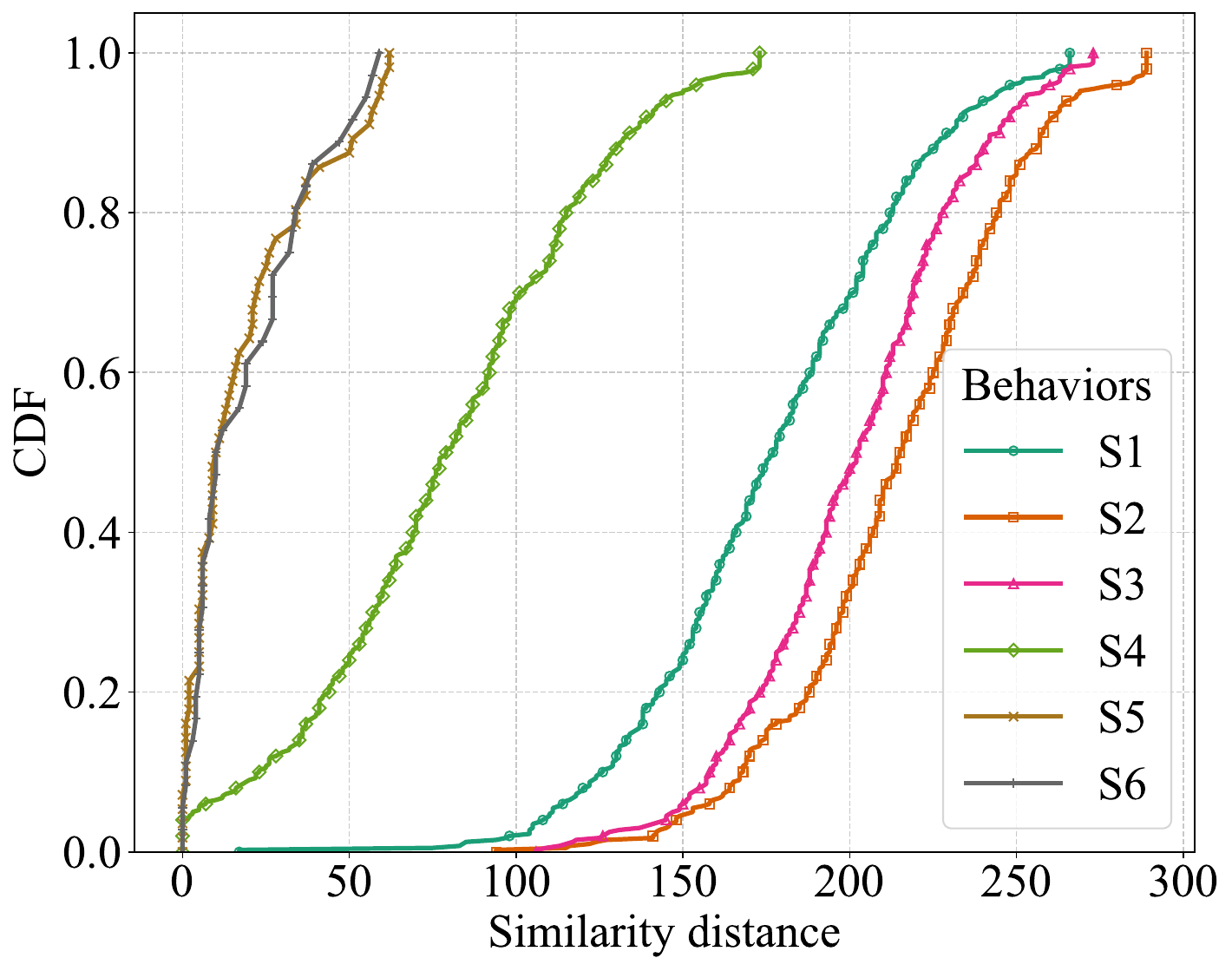}
    \caption{Distribution of similarity distances between inferred live behaviors of router pairs sharing the same behavior}
    \label{fig:cdf-router-similarity}
\end{figure}

To assess whether two I2P routers exhibit indistinguishable behavioral patterns, we analyze the similarity distance between the inferred live behaviors of controlled routers that share the same ground-truth behavior. For each day in the evaluation dataset(refer to \S\ref{subsec:exp_setup}), we compute the pairwise similarity distances among all controlled routers with identical ground-truth behaviors, as shown in \Cref{fig:cdf-router-similarity}. Routers with stable online patterns (\textbf{S5} and \textbf{S6}) display consistently low distances, with 85\% below 37 and a maximum of 62. In contrast, routers exhibiting frequent on–off transitions (\textbf{S1}--\textbf{S4}) show substantially larger distances ranging from 6 to 389. The results indicate that even when routers share the same ground-truth behavior, the similarity distance between their inferred live behaviors increases with the number of on–off switches. We therefore examined the distribution of online sessions in the real I2P network. Except for routers that remain continuously online, 99.74\% of routers experience fewer than four online sessions per day. Motivated by this observation, we recomputed similarity distances for controlled routers sharing the same behavior and exhibiting one to four sessions. Nearly all distances fall below 166, with only rare outliers reaching 189. Based on these results, we set the similarity threshold to 200, which provides a conservative upper bound ensuring that routers following identical behaviors are unlikely to be incorrectly excluded from each other’s anonymity sets, thereby minimizing false negatives. Although such a relaxed threshold may allow inclusion of some dissimilar routers, behavioral divergence that accumulates over longer observation windows can still enable an adversary to gradually shrink the target’s anonymity set.

\section{Implementation of I2Perception}
\label{subsec:implemetation}
We implemented a semi-automated prototype of the \textit{I2Perception} framework using the Python language. First, the system automatically pulls the previous day’s collected RouterInfo entries from the floodfill routers each day. Then, it autonomously infers each router's live behavior and completes the serialization task. Simultaneously, the sniffing client periodically accesses the target hidden service, using the sniffing results to infer the live behavior of the hidden service and serializes this information as well. Finally, the scheme automatically compares all the serialized live behaviors and provides the deanonymization results for the current monitoring epoch, outputting the anonymity set of the hidden service.

\section{Live behavior cases with data loss and non-data-loss cases}
\FloatBarrier
\label{app:mislead_cases}

\Cref{tab:special_cases} provides a comprehensive enumeration of all live-behavior cases that challenge our inference framework. Specifically, it includes (i) cases where the coarse-grained session inference becomes inaccurate under no–data-loss conditions (Cases 1--5 and Case 13), and (ii) cases where the fine-grained inference method may misidentify join or leave events due to data loss (Cases 6--12 and Cases 14--16). Whereas the main text (\Cref{subsec:time_state} and \Cref{subsec:complement}) discusses these cases individually when explaining the design rationale of each live behavior inference component, this appendix consolidates them in one place to support systematic cross-case comparison.

Each row of \Cref{tab:special_cases} represents a distinct live behavior pattern and the resulting \textit{RouterInfo} publication pattern of that behavior in either Java- or C++-based routers, along with the root cause of the inference error, a schematic comparison of inferred versus actual live behaviors, and a concise description of how the inference error arises. The last 3 columns indicate which technical components of our framework (join behavior identification, leave behavior identification, and online session complement) can correctly handle the case. The table thus highlights the functional contribution of every technical component and clarifies why all components are necessary to achieve robust live behavior inference.

\begin{table*}[!htbp]
    \scriptsize{
        \centering
        \begin{threeparttable}
        \caption{The situations needs to be handled and the functional contributions of each technical component across varying situations.}
        \begin{tabular}{M{0.6cm}| M{0.7cm}| M{0.7cm}| M{1.4cm}| M{2.6cm} M{3.3cm} M{1cm} M{1cm} M{1cm} M{1cm}}
        \toprule[1.5pt]
        \textbf{Index} & \textbf{Lang.} & \textbf{Data Loss} & \textbf{Root Cause} & \textbf{Schematic Diagram} & \textbf{Description} & \textbf{Coarse-Grained} & \textbf{+ Join Behavior} & \textbf{+ Leave Behavior} & \textbf{+ Session Complement} \\
        \midrule[1.5pt]
        
        1 & \multirow{12}{*}{\centering \rule[-2mm]{0mm}{57mm} \textbf{Java}} & \multirow{5}{*}{\centering \rule[-2mm]{0mm}{25mm} \ding{55}} & \multirow{2}{*}{\parbox{1.4cm}{\centering \rule[0mm]{0mm}{5mm} \textbf{Routine \textit{RouterInfo} mismatch}}} & \includegraphics[width=2.5cm]{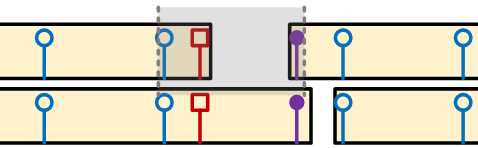} &  
        $I(R^i_r, R^{i+1}_f) = I_r$
        & \ding{55} & \ding{51} & - & - \\

        \cmidrule(lr){1-1} \cmidrule(lr){5-10}
        2 & & & & \includegraphics[width=2.5cm]{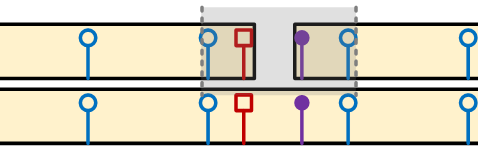} &  
        $I(R^i_r, R^{i+1}_r) = I_r$
        & \ding{55} & \ding{51} & - & - \\

        \cmidrule(lr){1-1} \cmidrule(lr){4-10}
        3 & & & \multirow{2}{*}{\parbox{1.4cm}{\centering \rule[0mm]{0mm}{2.7mm} \textbf{Routine \textit{RouterInfo} misidentification}}} & \includegraphics[width=2.5cm]{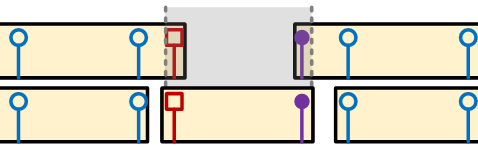} & 
        $I(R^i_{\ell}, R^{i+1}_f) = I_r$
        & \ding{55} & \ding{51} & - & - \\

        \cmidrule(lr){1-1} \cmidrule(lr){5-10}
        4 & & & & \includegraphics[width=2.5cm]{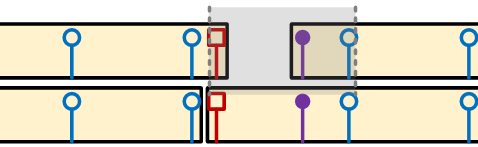} &  
        $I(R^i_{\ell}, R^{i+1}_r) = I_r$
        & \ding{55} & \ding{51} & - & - \\
        
        \cmidrule(lr){1-1} \cmidrule(lr){4-10}
        5 & & & \textbf{No routine feature} & \includegraphics[width=2.5cm]{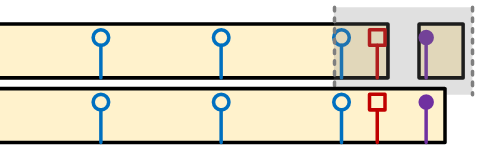} & 
        No routine \textit{RouterInfo} is published within the $(i+1)$\textsuperscript {th} online session. $I(R^i_r, R^{i+1}_f) < I_r$
        & \ding{55} & \ding{55} & \ding{51} & - \\

        \cmidrule(lr){1-1} \cmidrule(lr){3-10}
        6 & & \multirow{7}{*}{\centering \rule[-2mm]{-1mm}{34mm} \ding{51}} & \multirow{3}{*}{\parbox{1.4cm}{\centering \rule[0mm]{0mm}{10mm} \textbf{Loss of a routine \textit{RouterInfo}}}} & \includegraphics[width=2.5cm]{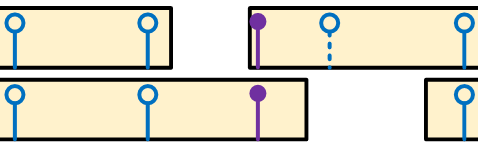} &  
        The first $R^{i+1}_r$ is lost. (non-floodfill only).
        & \ding{55} & \ding{55} & \ding{55} & \ding{51} \\

        \cmidrule(lr){1-1} \cmidrule(lr){5-10}
        7 & & & & \includegraphics[width=2.5cm]{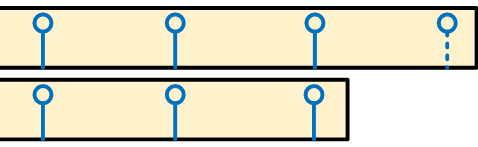} & 
        The last $R^{i}_{r}$ is lost. (non-floodfill only)
        & \ding{55} & \ding{55} & \ding{55} & \ding{55} \\
        
        \cmidrule(lr){1-1} \cmidrule(lr){5-10}
        8 & & & &  \includegraphics[width=2.5cm]{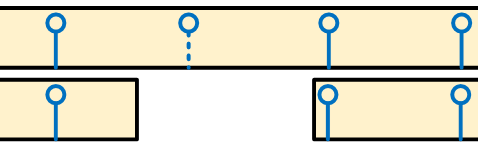} & 
        Loss of any $R^i_r$ other than the first or the last.
        & \ding{55} & \ding{55} & \ding{55} & \ding{51} \\
        
        \cmidrule(lr){1-1} \cmidrule(lr){4-10}
        9 & & & \textbf{Loss of the final \textit{RouterInfo}} & \includegraphics[width=2.5cm]{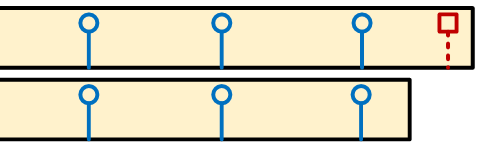} & 
        The $R^i_f$ is lost. Several $R^i_r$ may also lost. (floodfill only).
        & \ding{55} & \ding{55} & \ding{55} & \ding{51} \\

        \cmidrule(lr){1-1} \cmidrule(lr){4-10}
        10 & & & \multirow{3}{*}{\parbox{1.4cm}{\centering \rule[0mm]{0mm}{10mm} \textbf{Loss of the initial \textit{RouterInfo}}}} & \includegraphics[width=2.5cm]{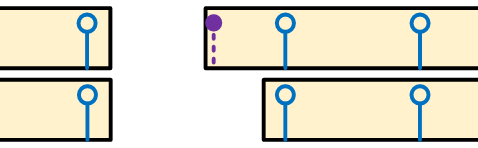} & 
        $R^{i+1}_f$ is lost, and $I(R^i_r, R^{i+1}_r) \neq I_r$.
        & \ding{55} & \ding{55} & \ding{55} & \ding{51} \\

        \cmidrule(lr){1-1} \cmidrule(lr){5-10}
        11 & & & & \includegraphics[width=2.5cm]{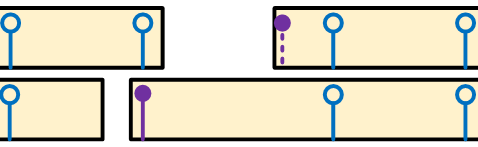} & 
        $R^{i+1}_f$ is lost, be mistakenly handled as Case 6 (non-floodfill only).
        & \ding{55} & \ding{55} & \ding{55} & \ding{55} \\

        \cmidrule(lr){1-1} \cmidrule(lr){5-10}
        12 & & & & \includegraphics[width=2.5cm]{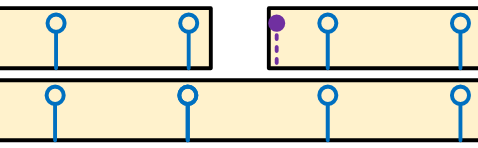} & 
        $R^{i+1}_f$ is lost, and $I(R^i_r, R^{i+1}_r) = I_r$, (non-floodfill only).
        & \ding{55} & \ding{55} & \ding{55} & \ding{55} \\
        
        \cmidrule(lr){1-10}
        13 & \multirow{4}{*}{\centering \rule[-2mm]{-1mm}{20mm} \textbf{C++}} & \ding{55} & \textbf{No routine feature} & \includegraphics[width=2.5cm]{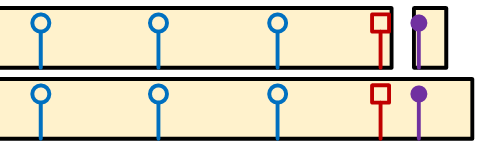} & 
        Identical to case 5.
        & \ding{55} & \ding{55} & \ding{51} & - \\
        
        \cmidrule(lr){1-1} \cmidrule(lr){3-10}
        14 & & \multirow{3}{*}{\centering \rule[-2mm]{-1mm}{15mm} \ding{51}} & \textbf{Loss of routine \textit{RouterInfo}} & \includegraphics[width=2.5cm]{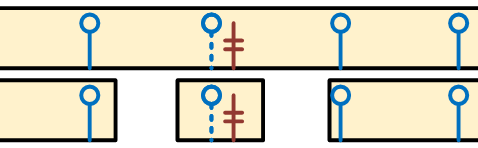} & 
        A $R^{i}_r$ is lost, the next \textit{RouterInfo} has different congestion flag with the previous $R^{i}_r$. 
        & \ding{55} & \ding{55} & \ding{55} & \ding{51} \\

        \cmidrule(lr){1-1} \cmidrule(lr){4-10}
        15 & & & \textbf{Loss of the initial \textit{RouterInfo}} & \includegraphics[width=2.5cm]{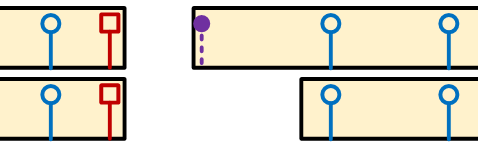} &  
        $R_f^{i+1}$ is lost.
        & \ding{55} & \ding{55} & \ding{55} & \ding{51} \\

        \cmidrule(lr){1-1} \cmidrule(lr){4-10}
        16 & & & \textbf{Loss of the final \textit{RouterInfo}} & \includegraphics[width=2.5cm]{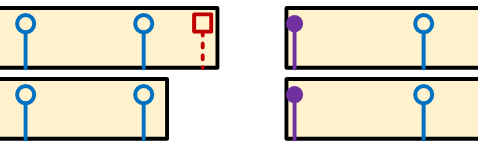} &  
        $R_{\ell}^{i+1}$ is lost.
        & \ding{55} & \ding{55} & \ding{55} & \ding{51} \\

        \bottomrule[1.5pt]
    \end{tabular}
    
    \label{tab:special_cases}
    \begin{tablenotes}\footnotesize
        \item \textbf{Diagram explanation}: Each diagram displays both the actual live behavior of the router (top section) and the inferred live behavior (bottom section). Each yellow rectangles represent router's one online session. Within each online session, the blue short line ending with a circle indicates the publish time of routine \textit{RouterInfo} data, the red short line with a hollow square marks the publish time of the final \textit{RouterInfo} data, and the purple short line terminating in a solid circle denotes the publish time of initial   \textit{RouterInfo} data within the session. Dashed line represents the \textit{RouterInfo} data that is lost. Additionally, in cases 1-5, gray rectangles between two online sessions indicate approximately one routine interval. Specifically, the time interval between the two \textit{RouterInfo} data located at the leftmost and rightmost edges of the gray rectangle equals one routine interval. 
        \item \textbf{Formal definition}: We define the function $I(\textit{RouterInfo 1}, \textit{RouterInfo 2})$ to calculate the time interval between two \textit{RouterInfo} data, with $I_r$ representing the routine interval. Next, we assign identifiers to important \textit{RouterInfo} data. For a given \textit{RouterInfo R}, the superscript (e.g., \textit{i}) indicates \textit{RouterInfo} data published within the $i$\textsuperscript {th} online session. Subscripts (e.g., \textit{r, f} and $\ell$) to indicate whether the \textit{RouterInfo} is a routine, the initial or the last \textit{RouterInfo} data within an online session, respectively.
    \end{tablenotes}
    \end{threeparttable}
    }
\end{table*}

\end{document}